\documentclass[useAMS,usenatbib]{mn2e}
\usepackage{times}
\usepackage{graphicx}
\usepackage{amssymb}
\usepackage{amsmath}
\usepackage{lscape}
\usepackage{longtable}

\title[SMBH - Galaxy Co-Evolution]{On the co-evolution of supermassive black holes and their host galaxies since z = 3}
\author[Asa F. L. Bluck et al.]{Asa~F.~L.~Bluck$^1$, Christopher~J.~Conselice$^1$,
Omar~Almaini$^1$, Elise~S.~Laird$^2$ \newauthor Kirpal~Nandra$^2$ and Ruth~Gr\"{u}tzbauch$^1$ \footnotemark[0]\\
\\$^1$ Centre for Astronomy and Particle Theory, School of Physics and Astronomy, University of Nottingham, Nottingham NG7 2RD, UK
\\$^2$ Astrophysics Group, Imperial College London, Blackett Laboratory, Prince Consort Road, London SW7 2AZ, UK }

\begin{document}

\maketitle

\begin{abstract}

To investigate the evolution in the relation between galaxy stellar and central black hole mass we identify a population of 508 X-ray selected AGN (Active Galactic Nuclei) at 0.4 $<$ z $<$ 6 residing within host galaxies with stellar masses in the range $10^{10} M_{\odot} < M_{*} < 10^{12} M_{\odot}$. From this sample we construct a volume limited complete sample of 85 AGN with host galaxy stellar masses $M_{*} > 10^{10.5} M_{\odot}$, and specific X-ray luminosities $L_{X} > 2.35 \times 10^{43}$ erg s$^{-1}$ at 0.4 $<$ z $<$ 3. We calculate the Eddington limiting masses of the supermassive black holes (SMBHs) residing at the centre of these galaxies, and observe an increase in the average Eddington limiting black hole mass with redshift. While the black hole mass and Eddington ratio, $\mu$, are degenerate, if we assume that the local $M_{BH} - M_{*}$ relation holds at all redshifts we find that the mean Eddington ratio $\mu$ rises from 0.056 +/- 0.010 at z = 0.7 to 0.087 +/- 0.011 at z = 1.25, with no significant evolution thereafter to z = 3.  Alternatively, by assuming that there is no evolution in $\mu$ and then that there is maximum possible evolution to the Eddington limit, we quantify the maximum possible evolution in the $M_{*} / M_{BH}$ ratio as lying in the range $700 < M_{*}/M_{BH} < 10000$, compared with the local value of $M_{*}/M_{BH} \sim 1000$.  We furthermore find that the fraction of galaxies which are AGN (with $L_{X} > 2.35 \times 10^{43}$ erg s$^{-1}$) rises with redshift from 1.2 +/- 0.2 \% at z = 0.7 to 7.4 +/- 2.0 \% at z = 2.5.  We use our results to calculate the maximum timescales for which our sample of AGN can continue to accrete at their observed rates before surpassing the local galaxy-black hole mass relation. We use these timescales to calculate the total fraction of massive galaxies which will be active (with $L_{X} > 2.35 \times 10^{43}$ erg s$^{-1}$) since z = 3, finding that at least $\sim$ 40\% of all massive galaxies will be Seyfert luminosity AGN or brighter during this epoch. Further, we calculate the energy density due to AGN activity in the Universe as 1.0 (+/- 0.3) $\times$ 10$^{57}$ erg Mpc$^{-3}$ Gyr$^{-1}$, potentially providing a significant source of energy for AGN feedback on star formation. We also use this method to compute the evolution in the X-ray luminosity density of AGN with redshift, finding that massive galaxy Seyfert luminosity AGN are the dominant source of X-ray emission in the Universe at $z < 3$.

\end{abstract}


\section{Introduction}

One of the major unresolved questions in observational astrophysics is the role of supermassive black holes (SMBHs) in the evolution of galaxies. For over a decade the existence of SMBHs in the centre of massive galaxies has been established (e.g. Kormendy \& Richstone 1995), but the impact they have on their host galaxies is still hotly debated. In the local Universe relations between the mass of the SMBH and the luminosity of the host galaxy have been discovered (Kormendy \& Richstone 1995), as well as tighter relations between the SMBH mass and velocity dispersion, and hence mass, of the hosting bulge or spheroid (Ferrarese \& Merritt 2000, Marconi \& Hunt 2003, Haring \& Rix 2004). These relationships indicate a causal link between galaxies and the SMBHs that reside within them. However, the nature of this interdependence and its origin remain open issues in the field. A classic crucial question is: which evolves first, the host galaxy or its SMBH? This fundamental `chicken and egg' question can be resolved by measuring the masses of SMBHs at higher redshifts and investigating whether they follow the local $M_{BH} - M_{*}$ relationship, and if not, quantifying how they differ from it.

SMBHs are thought to be a near ubiquitous component of massive galaxies (e.g. Ueda et al. 2003, Barger et al. 2005) although the exact fraction of massive galaxies that contain SMBHs is still unknown, especially at higher redshifts. One probe of the evolution of SMBHs is to look at the X-ray luminosity function (XLF) and examine  how it evolves with redshift (Ueda et al. 2003, Hasinger et al. 2005, Barger et al. 2005, Aird et al. 2009). These studies demonstrate that there is a steep positive evolution of the XLF with redshift out to z $\sim$ 1.2, followed by a less steep decline with redshift thereafter at even higher z. Studies disagree on whether there is evolution in the fundamental shape of the XLF, but all agree that there is evolution in the X-ray luminosity density with redshift. As such, SMBHs are known to be dynamic objects, subject to effects from the evolution of their host galaxies. In order to probe the evolution of AGN over cosmic time one must be careful to compare populations that are fundamentally linked. With XLF studies one considers a range of X-ray luminosities and quantifies the evolution within this range, often ignoring the characteristics of the underlying host galaxy population. In this paper, we restrict our sample of AGN to those with optical - NIR detected host galaxies with well determined redshifts and stellar masses, allowing us to link objects at different redshifts for comparison via the stellar masses of their host galaxies.

Measuring the mass of a SMBH, however, is not trivial. Often the most robust methods involve a virial mass estimation technique (see e.g. McLure \& Dunlop 2002), whereby one acquires detailed spectroscopy of the central most region of a galaxy and use the assumption of a virialised system to estimate the SMBH mass from the broadening of spectral lines. The mass contained within a radius $R$ is $M(<R) \sim R\sigma{^2} / G$. The value of $\sigma$ can be deduced directly from the FWHM (full width half maximum) of any broad line emission (most commonly $H_{\beta}$). More challengingly, the radius at which the broad line emitting region resides must be known in this method. This can be deduced from the time delay between line emission from the accretion disc and re-emission from the broad line emitting region, in orbit around the disc. Thus, $R \sim c\tau$, where $\tau$ is the time delay. This is only possible to measure in nearby systems with any accuracy, due to the high resolution required in this method. Fortuitously, an empirical relationship has been demonstrated to exist (e.g. McLure \& Jarvis 2002, Willott, McLure \& Jarvis 2003, Woo et al. 2008) between the monochrome luminosity of the 3000 \AA$\,$ line and the radius of the broad line emitting region. Nonetheless, even this method requires good spatial resolution, and a bright (most often quasar) source for anything but the most local AGN (active galactic nuclei).

In order to estimate the SMBH mass of a distant Seyfert-luminosity active galaxy, one may employ Eddington arguments by setting the outward radiative force equal to the inward force of gravity. Essentially this method gives a minimum value to the black hole mass given the luminosity of its accretion disc (see e.g. Alexander et al. 2009). Essentially $M_{BH} \propto \mu L_{Bol}$, with $\mu$ being a constant term related to the efficiency of the SMBH. In the local Universe, at z $<$ 0.2, $\mu \sim 0.05$ (see e.g. Marconi et al. 2009), implying that the average SMBH in the local Universe is accreting matter at $\sim$ 5 \% of the Eddington maximum. The bolometric luminosity of the accretion disc may be calculated by employing a measured X-ray specific luminosity and utilising a spectral energy distribution. Naturally, these Eddington based methods are less reliable for calculating the mass of the central black hole because they contain an efficiency degeneracy. Therefore, one can trace the evolution of the black hole mass only with assumptions about the evolution of the efficiency and vice versa. However, for representative AGN at intermediate to high redshifts (z $>$ 1) the Eddington method is often the only method for calculating SMBH masses for more than a few galaxies as it does not depend on obtaining very high resolution spectroscopy which frequently cannot be obtained, especially for lower luminosity AGN.

In this paper we calculate the Eddington limiting (minimum SMBH) mass for 508 X-ray active galaxies with known stellar masses and spectroscopic or photometric redshifts, taken from the GOODS and Extended Growth Strip fields. We utilise deep X-ray data from the Chandra X-ray Observatory (hereafter Chandra) and optical to near infrared data from the Hubble Space Telescope (HST) NICMOS and ACS cameras, the Palomar Observatory and a host of other ground based observatories, including the CFHT, VLT and Keck observatories. We construct from this data set a volume limited sample of 85 `Seyfert' galaxies with specific hard band luminosities $L_{X} > 2.35 \times 10^{43} {\rm erg s^{-1}}$ and host galaxy stellar masses in the range $10^{10.5} M_{\odot} < M_{*} < 10^{12} M_{\odot}$ (selected to be not point like), where we are complete to z = 3. We use this sample to probe the evolution of the $M_{BH} - M_{*}$ relationship with redshift, and to compute the total active fraction (at bright `Seyfert' luminosities) of massive galaxies over the past 11.5 billion years.

This paper is structured as follows: \S 2 outlines the sources of our data and measures of photometric redshifts and stellar masses, \S 3 outlines the major sources of random and systematic error or biases affecting this work, and what we have done to parameterise and minimise their effects. \S 4 outlines our methods, including explanation of how we deduce lower limits to black hole masses. \S 5 presents our results in detail, as well as presenting a novel method for calculating AGN lifetimes. \S 6 encompasses a discussion of our results, with \S 7 summarising our conclusions. Throughout the paper we assume a $\Lambda$CDM Cosmology with: H$_{0}$ = 70 km s$^{-1}$ Mpc$^{-1}$, $\Omega_{m}$ = 0.3, $\Omega_{\Lambda}$ = 0.7, and adopt AB magnitude units.

\section{Data and Observations}

We utilise near-infrared and optical data from the GOODS ACS and NICMOS galaxy Surveys (ACS, Dickinson et al. 2003; GNS, Conselice et al. 2010, in prep), the Palomar Observatory Wide-field Infra-Red (POWIR) galaxy Survey (Conselice et al. 2007, 2008, and Davis et al. 2007), and complementary surveys based in the GOODS field and Extended Growth Strip (EGS). Additionally, we compare these observations to X-ray catalogs from the Chandra Deep Field North / South (CDF-N/S, Alexander et al. (2003), Luo et al. 2008), and the AEGIS-X survey (Laird et al. 2009). In so doing we construct two samples: an `AGN' sample where we are not complete (with all NIR galaxy - X-ray source matches within a 1.5$''$ radius for hard band X-ray luminosities $L_{X} > 10^{42}$ erg s$^{-1}$ in galaxies with stellar masses $M_{*} > 10^{10} M_{\odot}$; and a volume limited sample of `Seyfert' galaxies (defined to be not-point-like AGN with hard band X-ray luminosities $L_{X} > 2.35 \times 10^{43}$ erg s$^{-1}$ in host galaxies with sellar masses $M_{*} > 10^{10.5} M_{\odot}$) where we achieve completion to z = 1.5 in the EGS field and z = 3 in the GNS fields. In general we use the first, and larger, sample to probe intrinsic AGN properties, and the volume limited sample to probe the redshift co-evolution of SMBHs and their host galaxies in the `Seyfert' luminosity regime.

\subsection{Near Infrared Data}

We utilise the POWIR survey in the EGS field to obtain redshifts and stellar masses for a large sample of galaxies and, after matching with Chandra data, our sample of AGN hosts out to redshifts of z = 1.5. The POWIR Survey obtained deep K and J band imaging over $\sim$ 1.52 deg$^{2}$ in the EGS (see Fig. 1 for an illustration). Limiting magnitudes reached in the K band are typically 22.5 - 23 (5 $\sigma$) over the entire area. In total $\sim$ 20,000 galaxies were imaged between z = 0.4 and z = 2. Photometric redshifts were calculated for these galaxies using optical (CFHT) + NIR (Palomar) imaging in the BRIJK bands (see Conselice et al. 2008 for a detailed discussion of these, and \S 2.2 for further details). From this stellar masses were estimated by fitting spectral energy distributions with Bruzual \& Charlot (2003) stellar population synthesis models, with varying star formation histories. Resulting stellar mass errors are $\sim$ 0.2 - 0.3 dex (see Conselice et al. 2007 and \S 2.3 for further details). We are complete down to $M_{*} = 10^{10.5} M_{\odot}$ at z = 1.5. For full details on the selection, photometry, redshift and stellar mass calculations see Conselice et al. (2007 and 2008) and \S 2.2 and 2.3. Also see \S 3.2 for a discussion of possible sources of systematic errors on photometric redshifts and stellar masses.

We utilise data from the ultra-deep `pencil beam' HST GOODS ACS and GNS surveys to obtain masses and photometric redshifts for galaxies at high redshifts (1.5 $<$ z $<$ 3) in the GOODS North and South fields (see Fig. 2). We use published masses and photometric redshifts (see Dickinson et al. 2003, Giavalisco et al. 2004)) for the large AGN sample, but restrict for our volume limited sample the area probed to that covered additionally by the GNS, where we have much deeper imaging in the H-band and higher mass completion. The GOODS NICMOS Survey images a total of $\sim$ 8000 galaxies in the F160W (H) band, utilising 180 orbits and 60 pointings of the HST NICMOS-3 camera (see Fig. 2 for a depiction of the pointings). These pointings are centred around massive galaxies at 1.7 $<$ z $<$ 3 to 3 orbits depth. Each tile (51.2$''$x51.2$''$, 0.203$''$/pix) is observed in 6 exposures that combine to form images with a pixel scale of 0.1$''$, and a point spread function (PSF) of $\sim$ 0.3$''$ full width half maximum (FWHM). The total area of the survey is $\sim$ 43.7 arcmin$^{2}$. Limiting magnitudes reached are H = 26.5 (10$\sigma$). We are complete down to $M_{*} = 10^{10} M_{\odot}$ at z = 3, to 10 $\sigma$. The wealth of observational data available in the GOODS fields allows us to utilise data from the U to the H band in the determination of photometric redshifts (see \S 2.2 for further details). Multi-colour stellar population fitting techniques were utilised to estimate stellar masses to an accuracy of $\sim$ 0.2 - 0.3 dex (see \S 2.3 for further details and \S 3.2 - 3.2 for a discussion on the errors involved in computing stellar masses for AGN at high redshifts). A selection bias is introduced due to our pointings being selected to maximise the number of massive galaxies imaged, whereby we witness higher surface densities of massive galaxies in the GNS than expected from imaging to a similar depth a randomly chosen area of sky. Where necessary we take account of this bias in our analyses (see e.g. \S 5.4.5).

Details of the techniques used for the GNS in constructing catalogs, photometric redshifts and stellar masses may be found in Buitrago et al. (2008), Bluck et al. (2009), Gruetzbauch et al. (2010) and a more full account in Conselice et al. (2010), in prep., as well as further information in \S 2.2 - 2.3. A discussion of more robust error estimates on stellar masses can be found in Bluck et al. (2009), where it is concluded, through Monte-Carlo simulation, that systematic errors arising out of Poisson errors from the steepness of the stellar mass distribution (Eddington biases) do not result in a significant infiltration of lower mass objects. Further discussion on the use of the SExtractor package to produce galaxy catalogs is also provided in Bluck et al. (2009), with special concern for deblending accuracy and removal of stellar objects and spurious detections. Possible sources of systematic bias on stellar masses and photometric redshifts, and their effect on our results, are addressed in detail in \S 3.

\subsection{Spectroscopic and Photometric Redshifts}

We utilise both spectroscopic and photometric redshifts for the
galaxies we study in both the EGS and GNS fields. The only
field which has extensive available spectroscopy, however,
is the EGS. The Keck EGS spectra were acquired with the DEIMOS spectrograph 
as part of the DEEP2 redshift survey (Davis
et al. 2003). Target selection for the DEEP2 spectroscopy
was based on the optical properties of the galaxies detected in the
CFHT photometry, with the basic selection criteria being $R_{\rm AB} < 24.1$.   
DEEP2 spectroscopy was acquired through
this magnitude limit, with no strong colour cuts applied to the selection.
About 10,000 redshifts are measured for galaxies within the EGS. The 
sampling rate for galaxies that meet the selection criteria is
60\%.

This DEIMOS spectroscopy was obtained using the
1200 line/mm grating, with a resolution R $\sim 5000$ covering
the wavelength range 6500 - 9100 \AA.  Redshifts were measured through
an automatic method comparing templates to data, and we only utilise
those redshifts measured when two or more lines were
identified, providing very secure measurements. Roughly 70\% of all
targeted objects result in secure redshifts.

We utilise photometric redshifts computed by our group within the EGS (e.g. 
Bundy et al. 2006; Conselice et al. 2007, 2008). Within the EGS, photometric 
redshifts are based on the optical + near infrared imaging, in the BRIJK  bands, 
and are fit in two ways, 
depending on the brightness of a galaxy in the optical. For galaxies that
meet the spectroscopic criteria, $R_{\rm AB} < 24.1$, we utilise a neural
network photometric redshift technique to take advantage of the
vast number of secure redshifts with similar photometric data.  Most
of the $R_{\rm AB} < 24.1$ sources not targeted for spectroscopy should be 
within our redshift range of interest, at $z < 1.5$.    

The neural network fitting is done through 
the use of the ANNz (Collister \& Lahav 2004) method and code.
To train the code, we use the $\sim 5000$ secure redshifts in the EGS, which
have galaxies spanning our entire redshift range. The training of the 
photometric redshift fitting was in fact only done using the EGS field, whose
galaxies are nearly completely selected based on a magnitude
limit of $R_{\rm AB} < 24.1$. We then use this training to calculate the
photometric redshifts for galaxies with $R_{\rm AB} < 24.1$.   
The agreement between our photometric redshifts and our ANNz 
spectroscopic redshifts is very good 
using this technique, with $\delta z/(1+z) = 0.07$ out
to $z \sim 1.4$. The agreement is even better for the $M_{*} > 10^{10.5} M_{\odot}$
galaxies where we find $\delta z/(1+z) = 0.025$ across all of our
four fields (Conselice et al. 2007).

For galaxies which are fainter than $R_{\rm AB} = 24.1$ in the EGS
we utilise photometric redshifts using Bayesian techniques, and the 
software from Benitez (2000). For an object to have a photometric redshift
we require that it be detected at the 3 $\sigma$ level in all
optical and near-infrared (BRIJK) bands, which in the R-band
reaches $R_{\rm AB} \sim 25.1$. We 
optimised our results, and correct for systematics, through 
the comparison with spectroscopic redshifts, resulting
in a redshift accuracy of $\delta z/(1+z) = 0.17$ for $R_{\rm AB} > 24.1$ systems. 
These $R_{\rm AB} > 24.1$ galaxies are, however, only a very small part of our
sample.  Furthermore, all of these systems are at $z > 1$.    

The redshifts for the GOODS sample were derived using the HST ACS and ground based
NIR imaging, for the whole GOODS area, and the HST GNS + ACS space based data for the volume limited sub-sample. Most of these sources were chosen through multiple selection
methods, namely the BzK, IERO and DRG selection (see Conselice et al. 2010, in prep. for
a full description). However, in the volume limited case only mass selection was applied (see Fig. 2). Our photometric redshifts, which
we use for these massive galaxies, are determined via standard techniques similar
to that used in the POWIR data described above (see Gruetzbauch et al. 2010, Conselice et al. 2010 in prep.).
Additionally, we find seven spectroscopic redshifts 
from the literature for our sample of massive galaxies. Using the GOODS/VIMOS DR1 (see Popesso et al. 2008), we find three matches with $\delta z/(1+z) = 0.026$, and four spectroscopic redshifts from a compilation of redshifts from the literature 
(see Wuyts et al. 2008) giving  $\delta z/(1+z) = 0.034$. Please see Conselice et al. 2010, in prep., for full details on the determination of photometric redshifts and stellar masses for the GNS sample.

\subsection{Stellar Masses}

For the POWIR survey in the EGS field, stellar masses were derived from spectroscopic and photometric redshifts and Palomar Obaservatory and Canada France Hawaii Telescope obtained BRIJK wavebands by fitting spectral energy distributions with Bruzual \& Charlot (2003) stellar population synthesis models, with varying star formation histories. A Chabrier initial mass function (IMF) was assumed. Full details of this method and its intrinsic errors are provided in Conselice et al. (2007 and 2008). The errors on the stellar masses were estimated to be $\sim$ 0.2 - 0.3 dex, taking into account issues regarding the reliability of photometric redshifts, the effects of AGB stars on the analysis, and Eddington bias. Additional errors from the choice of IMF for the stellar mass codes would lead to a factor of two or so uncertainty (see Conselice 2007). 

For the full GOODS field (non-volume limited sample), stellar masses were derived from a combination of groundbased NIR imaging, and space based (ACS) optical imaging. These were taken from the literature (see Giavalisco et al. 2004).

For the GNS, volume limited, sample stellar masses were calculated from photometric redshifts and the HST ACS (BViZ) and NICMOS (H) wavebands. The exact same Bruzual \& Charlot (2003) models and Chabrier IMF were used as with the POWIR sample. Full details of the techniques used and error analyses performed are provided in Conselice et al. (2010), in prep. and further details in Gruetzbauch et al. (2010). Errors on these stellar masses are also estimated to be $\sim$ 0.3 dex, with possible additional errors on top from the choice of IMF.

The effect of possible systematic bias on the determination of stellar masses for high redshift AGN host galaxies is discussed in detail in \S 3.1 and \S 3.2. Here systematic errors arising from the contamination of AGN light in the galaxy SED, and from systematic photometric redshift errors (and catastrophic outliers) for high redshift AGN host galaxies are considered.

\subsection{X-ray Data}

The X-ray data used in this paper all originates from the Chandra X-ray Observatory, with pointings in the GOODS North / South and EGS fields (see Fig. 1 and 2 for a graphical representation of the fields). The AEGIS-X survey covers a $\sim$ 0.64 deg$^{2}$ area on the sky with on-axis limiting fluxes of 5.3 $\times$ 10$^{-17}$ erg cm$^{-2}$ s$^{-1}$ in the soft (0.5 - 2 KeV) band and 3.8 $\times$ 10$^{-16}$ erg cm$^{-2}$ s$^{-1}$ in the hard (2 - 10 KeV) band. The AEGIS-X survey is complete for 54 \% of the total survey area covered in the hard band down to 2.1 $\times$ 10$^{-15}$ erg cm$^{-2}$ s$^{-1}$. This corresponds to a X-ray luminosity of 2.35 $\times$ 10$^{43}$ erg s$^{-1}$ at z = 1.5, which is the redshift limit we use for the POWIR survey. Thus, we are complete in X-ray sources above our luminosity threshold for our volume limited sample, with $L_{X}$ $>$ 2.35 $\times$ 10$^{43}$ erg s$^{-1}$ at z $<$ 1.5, for a reduced area of the total EGS field. We restrict our volume limited sample to those galaxies and X-ray sources which reside within the fraction of the survey's area where we are complete to these limits. The total area of our volume limited sample is approximately 1/2 of the total AEGIS-X area (corresponding to $\sim$ 1/4 of the total EGS POWIR field area, see Fig. 1). See Laird et al. (2009) for further details on this survey.

The X-ray data we use to compare with the GOODS North galaxy catalogs comes from the Chandra Deep Field North 2 Ms X-ray source catalog (which along with the Chandra Deep Field South is the deepest ever X-ray image taken to date). This covers $\sim$ 460 arcmin$^{2}$ of sky and reaches on-axis soft band (0.5 - 2 KeV) fluxes of 2.5 $\times$ 10$^{-17}$ erg cm$^{-2}$ s$^{-1}$ and on-axis hard band (2 - 8 KeV) fluxes of 1.4 $\times$ 10$^{-16}$ erg cm$^{-2}$ s$^{-1}$. This survey is complete for 28.5 \% of the total survey area down to a hard band flux of 4.5 $\times$ 10$^{-16}$ erg cm$^{-2}$ s$^{-1}$, which corresponds to a luminosity limit of $L_{X}$ = 2.3 $\times$ 10$^{43}$ erg s$^{-1}$ at z = 3, where we limit our GOODS data in redshift for the volume limited sample. This leads us to detect all X-ray sources above our luminosity threshold ($L_{X}$ $>$ 2.35 $\times$ 10$^{43}$ erg s$^{-1}$) at z $<$ 3 within the reduced area, to which we restrict our volume limited sample. See Alexander et al. (2003) for a comprehensive review of the CDF-N data acquisition and properties.

To compare with our GOODS South galaxy catalogs we use the CDF-S 2 Ms X-ray source catalog. Here an area of $\sim$ 436 arcmin$^{2}$ of sky is imaged in X-ray hard and soft bands. Limiting on-axis fluxes achieved are: soft band (0.5 - 2 KeV) 1.9 $\times$ 10$^{-17}$ erg cm$^{-2}$ s$^{-1}$ and hard band (2 - 10 KeV) 1.3 $\times$ 10$^{-16}$ erg cm$^{-2}$ s$^{-1}$. This survey is complete for 22 \% of the total survey area down to fluxes of 4.5 $\times$ 10$^{-16}$ erg cm$^{-2}$ s$^{-1}$. This corresponds to $L_{X}$ = 2.3 $\times$ 10$^{43}$ erg s$^{-1}$ at z = 3, where we limit our sample. This leads us to detect all X-ray sources above our luminosity threshold ($L_{X}$ $>$ 2.35 $\times$ 10$^{43}$ erg s$^{-1}$) at z $<$ 3 within the reduced area, to which we restrict our volume limited sample. For more details relating to the CDF-S data see Luo et al. (2008).

Our volume limited sample within the highest redshift range, taken from the GOODS North and South fields, has an area of $\sim$ 2/3 that of the total GNS field, which corresponds to $\sim$ 1/9 of the area of the entire GOODS North and South fields. (See Table 1 for a summary of the data and the areas of completion).

Positional uncertainties are of the order 1 arcsec in size for all of the X-ray data used, making it possible to place a strict 1.5 arcsec aperture limit on confirming an X-ray detected galaxy through matching with NIR imaging (from the GNS and POWIR surveys). Additionally, we imposed checks to determine how sensitive our detection of AGN is to the exact aperture limits we use, finding very few additional detections ($\sim$ 5 \%) by expanding the limits up to 2.5 arcsecs, and very little loss ($\sim$ 2 \%) by reducing the limit to 1 arcsec. We also use the criterion of ensuring each detection is unique, to avoid possible confusion. In fact, no AGN - galaxy matches had to be rejected for this reason using an aperture of 1.5 arcsecs.

We present in Table 1 a summary of the key surveys utilised in this paper, which indicates their areas, depths and the reduced areas used to construct volume limited samples. In the table, $A_{T}$ is the total area of the survey(s) in question, $A_{vol. lim.}$ is the area of the volume limited sub-sample of the survey where there is completion to the mass limits and redshifts indicated (for the NIR + optical surveys), and to a threshold hard band X-ray luminosity of $L_{X}$ $>$ 2.35 $\times$ 10$^{43}$ erg s$^{-1}$ at the redshift limit indicated for each X-ray survey. $M_{L}$ is the mass limit completion cutoff for the volume limited sample. The quoted depths are given for hard band fluxes (in the case of the X-ray data) and magnitudes in the wavebands indicated (for the NIR + optical surveys). Please see above (\S 2) for full details on the specifics of each survey, and \S 4 for further details on the construction of our volume limited sample.

\vspace{1cm}
\setcounter{table}{0}
\begin{table*}
 \caption{Data Source Summary}
 \label{tab1}
 \begin{tabular}{@{}ccccccc}
  \hline
\hline
Field & Waveband(s) & $A_{Tot}$ [deg$^{2}$] & $A_{vol. lim.}$ [deg$^{2}$] & Depth & $M_{*Lim}$ [$M_{\odot}$] & z Range \\
\hline
EGS (POWIR + CFHT) & BRIJK & 1.5200 & 0.3466 & 23.0 (AB: K) & 10$^{10.5}$ & 0.4 - 1.5 \\
GOODS (GNS + ACS) & BViZH & 0.0120 & 0.0079 & 26.5 (AB: H) & 10$^{10}$ & 1.5 - 3.0 \\
AEGIS-X & 0.5-2, 2-10 KeV & 0.6419 & 0.3466 & 3.8 $\times$ 10$^{-16}$ erg cm$^{-2}$ s$^{-1}$ & .... & 0.4 - 1.5 \\
CDF-N & 0.5-2, 2-8 KeV & 0.1211 & 0.0044 & 2.5 $\times$ 10$^{-17}$ erg cm$^{-2}$ s$^{-1}$ & .... & 1.5 - 3.0 \\
CDF-S & 0.5-2, 2-8 KeV & 0.1212 & 0.0035 & 1.9 $\times$ 10$^{-17}$ erg cm$^{-2}$ s$^{-1}$ & .... & 1.5 - 3.0 \\
\hline
\end{tabular}
\end{table*}

\begin{figure}
\includegraphics[width=0.5\textwidth,height=0.5\textwidth]{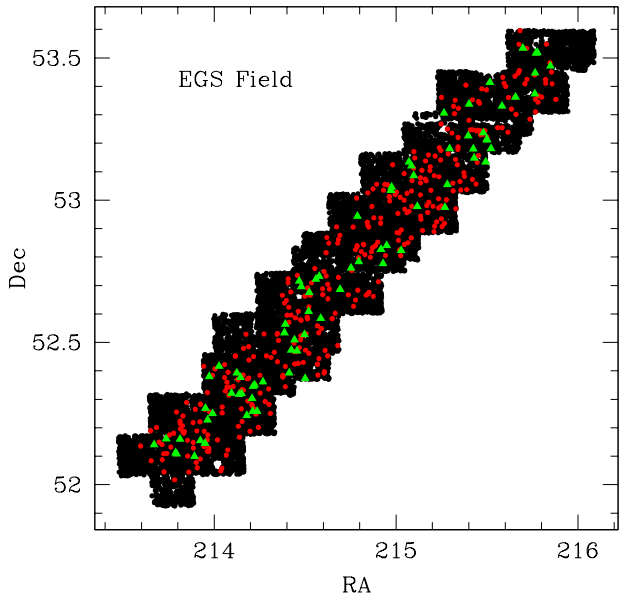}
\caption{The EGS field. Black dots/ region indicate the location of $M_{*} > 10^{10} M_{\odot}$ galaxies at redshifts (0.4 $<$ z $<$ 1.5). Note that there are $\sim$ 16000 galaxies displayed here. The red circles are massive galaxies with secure X-ray counterparts within a 1.5$''$ radius of their positional centre, with hard band luminosities $L_{X}$ $>$ 10$^{42}$ erg s$^{-1}$, which form our low and intermediate redshift AGN sample. The Chandra (AEGIS-X) survey covers an area approximately 1/2 the size of the POWIR survey, which can be visualised by looking for where the red points generally reside. The green triangles indicate bright `Seyfert' luminosity AGN with $L_{X}$ $>$ 2.35 $\times$ 10$^{43}$ erg s$^{-1}$ at 0.4 $<$ z $<$ 1.5, residing within massive galaxies with $M_{*} > 10^{10.5} M_{\odot}$, where we are complete for a reduced area of this survey ($\sim$ 1/4 the entire EGS area). The criteria to select the region of completion for the Chandra AEGIS-X survey is roughly given by setting the off-axis angle OAA $<$ 7.02$'$. This marks out a complex area due to varying depths, angles of pointing, and repeat observations, and hence we do not display it here. However, as a guide to the region of completion, where the green triangles reside we are complete to the above stated limits.}
\end{figure}

\begin{figure*}
\includegraphics[width=0.45\textwidth,height=0.45\textwidth]{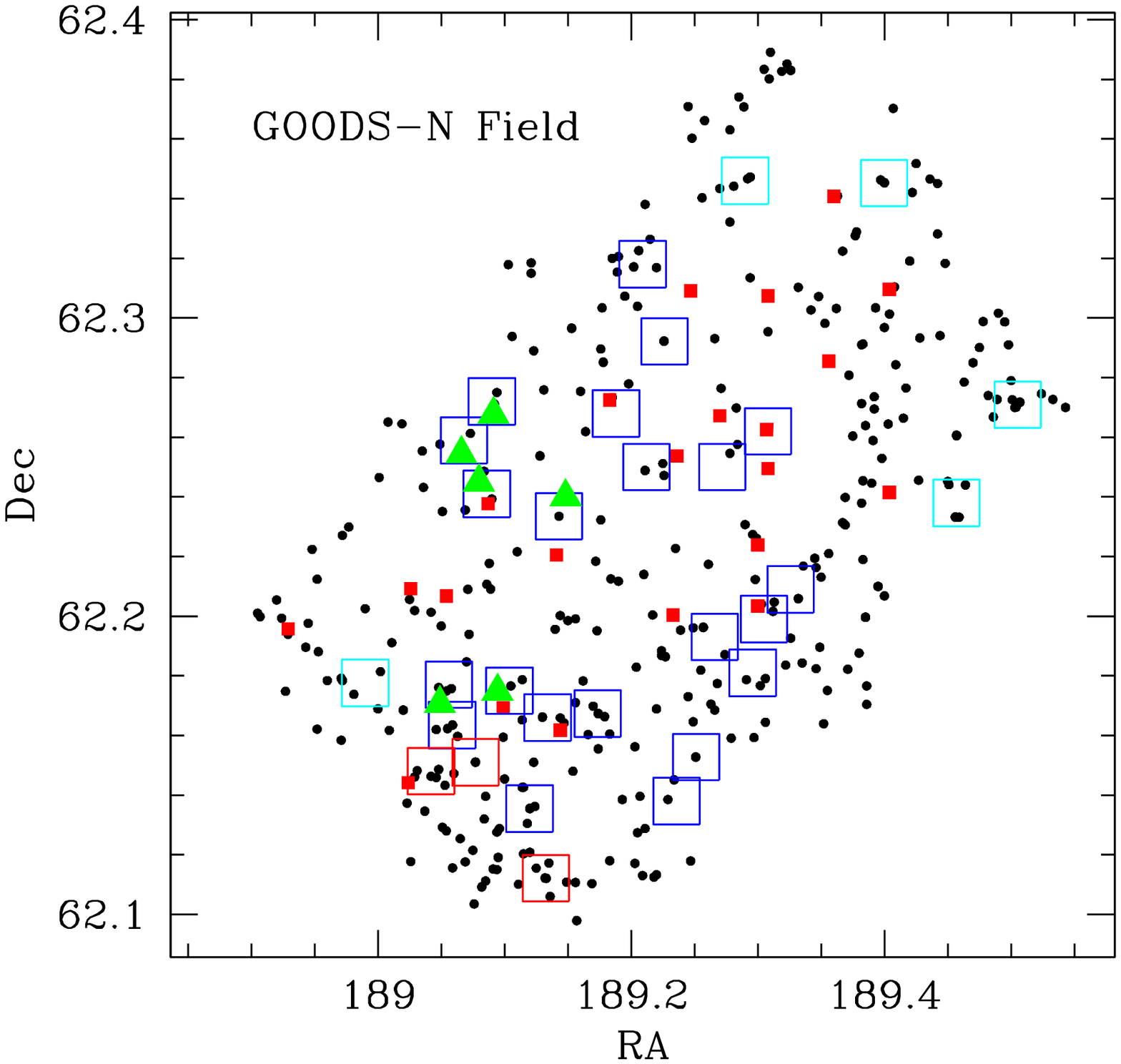}
\includegraphics[width=0.45\textwidth,height=0.45\textwidth]{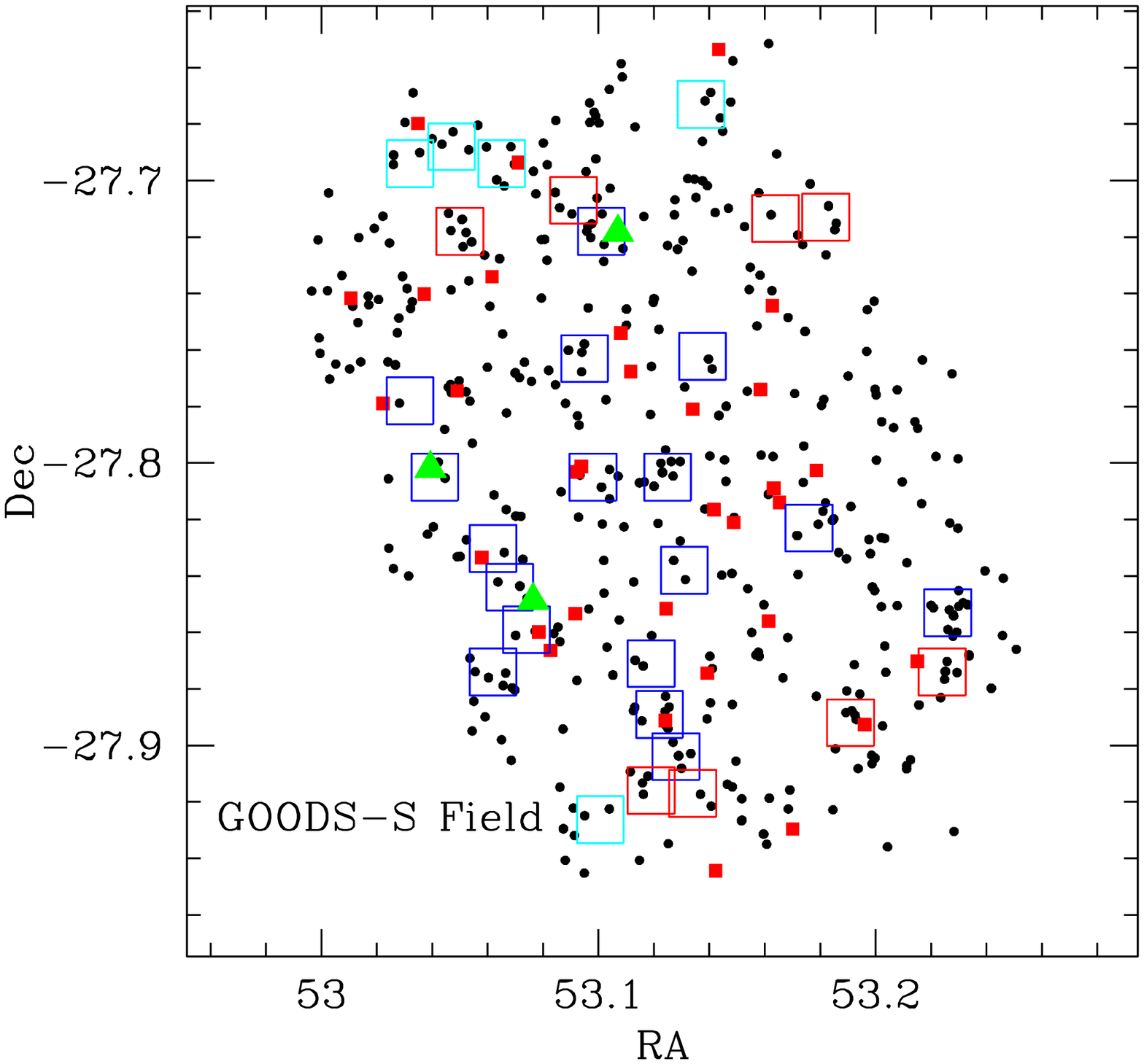}
\caption{Left plot is the GOODS North field; right plot is the GOODS South Field. The black dots indicate the location of $M_{*} > 10^{10} M_{\odot}$ galaxies at high redshifts (1.5 $<$ z $<$ 3) throughout the entire GOODS field. Small red sqaures depict those high redshift massive galaxies within the GOODS field with secure X-ray counterparts within a 1.5$''$ radius of their positional centre, with hard band X-ray luminosities $L_{X}$ $>$ 10$^{42}$ erg s$^{-1}$, which form our high redshift AGN sample. The large green triangles indicate bright `Seyfert' luminosity AGN with $L_{X}$ $>$ 2.35 $\times$ 10$^{43}$ erg s$^{-1}$ residing within massive galaxies with $M_{*} > 10^{10.5} M_{\odot}$, where we are complete for a reduced area of this survey ($\sim$ 1/9 the entire GOODS area). The sqare boxes represent the pointings of the HST GOODS NICMOS Survey (GNS), and these are colour coded via Chandra depth. The blue boxes have total Chandra completion to a depth of $L_{X}$ $>$ 2.35 $\times$ 10$^{43}$ erg s$^{-1}$ at z = 3 across their entire area, the cyan boxes have partial completion to this limit, whereas the red boxes are totally insensitive to this threshold. As such, our volume limited sample is restricted to the blue box regions (where all of our sample of `Seyfert' galaxies reside).}
\end{figure*}

\section{Biases and Systematic Errors}

Biases and systematic errors are likely an important aspect in our analysis and we have to
address these issues very carefully. Not only do we have to deal with issues such as
intrinsically inaccurate stellar masses and photometric redshifts (from random errors), but also, we have the added complication
that our sources are AGN and emit light in the optical and NIR which can make quantities such
as stellar masses and photometric redshifts systematically less accurate. Some of these systematics include contamination of
the AGN in creating higher stellar masses due to the possibly bright point source, as well as creating a
galaxy SED which is not entirely stellar. Another issue is that the Chandra X-ray 
Observatory data is not uniform across its field of view, resulting in less sensitive
detections in the outer regions of the field. 

We also may detect slightly fewer AGN above our flux limits due to the randomly distributed torus opening angles of accretion discs in our survey. In part this issue will be mitigated by the fact that we utilise the hardest X-ray bands (and hence most penetrating X-ray emission) for our study but this effect is still likely to make our measures of active fractions and luminosity densities lower limits as discussed throughout this paper. Studies in the local Universe, utilising data from the Swift and INTEGRAL surveys, still find very few Compton thick sources (i.e. those most likely to have edge on accretion discs to our line of sight) even up to energies of $\sim$ 200 KeV (see Ueda et al. 2003 and Steffen et al. 2003). This suggests that missing AGN due to certain torus opening angles will still be a significant issue even though we use the deepest and hardest available X-ray bands to probe our sources. Additionally, these surveys find that those highly obscured AGN have lower luminosities than their less obscured counterparts. This is, however, not a significant problem for the results and conclusions in this paper because we consider our active fractions and total energy outputs to be minima for this and other reasons, as is explained in detail in the relevant sections. Finally, to
derive an Eddington limited black hole mass, we must have some idea of the bolometric
luminosity of the AGN. We explore each of these issues in turn throughout this section.

\subsection{AGN Contamination in Stellar Masses}

Since we are working with galaxies that can host bright AGN, we must deal with the fact that some light in the optical-NIR spectral energy distributions of these galaxies may originate from light coming from the AGN. A few methods and tests were used to ensure that this is not a significant problem. The first is that our sample does not include any objects that are point sources, as these were removed before the analysis of stellar masses in both the POWIR data (Conselice et al. 2007; 2008) and the GNS samples (Conselice et al. 2010, in prep). Secondly, in terms of stellar masses, these were measured using only galaxy SEDs, and we find no cases for our sample where the SED is not well fit by standard star formation histories. Regardless, the stellar masses are measured in such a way that the uncertainty in the stellar masses takes into account the range of best fit masses, and this uncertainty increases if the SED is not well fit by a limited number of models, giving a wider range of calculated stellar masses for our library of star formation histories (see Bundy et al. 2006; Conselice et al. 2007; Bundy et al. 2008).

\begin{figure}
\includegraphics[width=0.5\textwidth,height=0.5\textwidth]{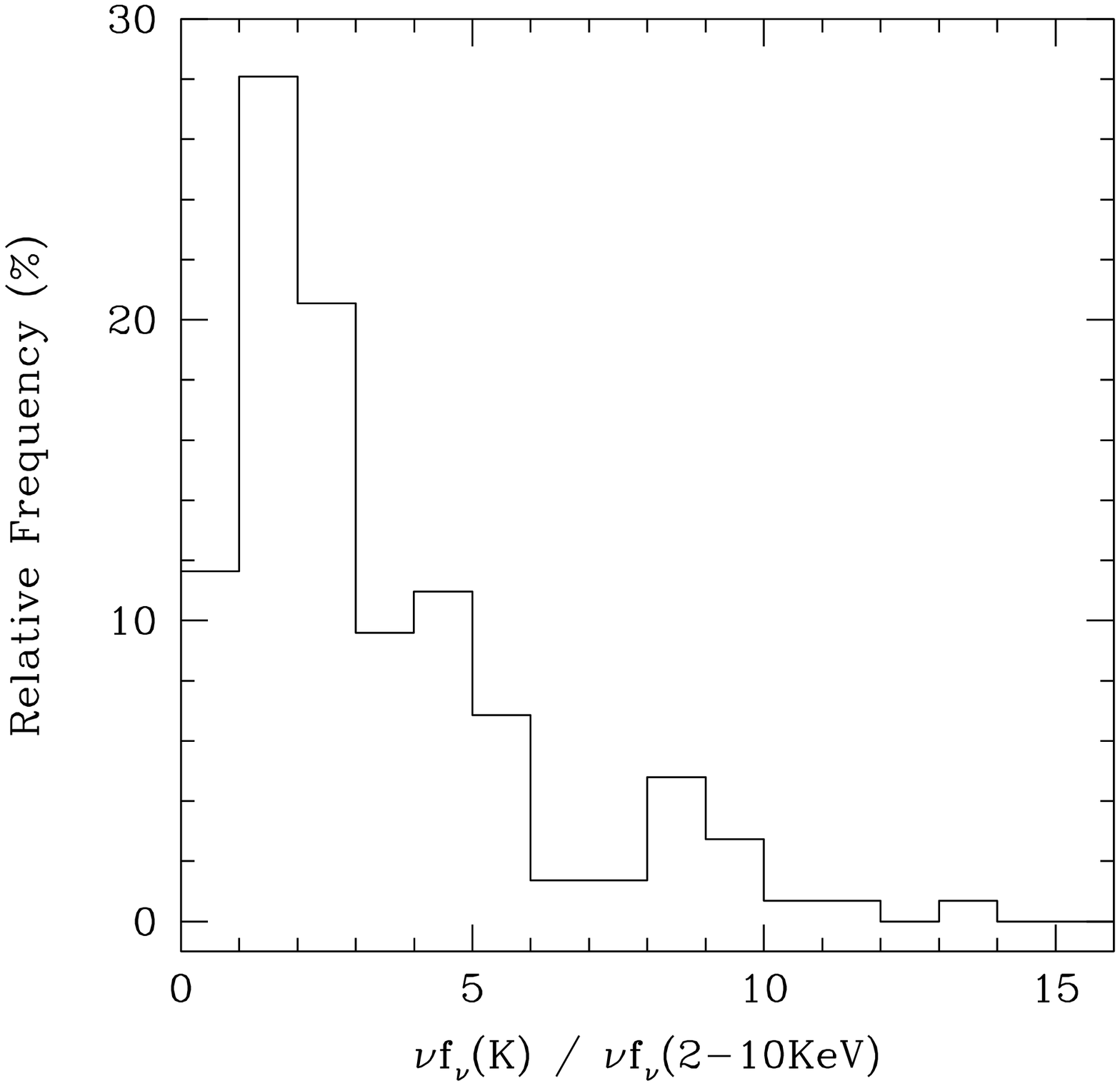}
\caption{Frequency histogram of the ratio of K band flux to X-ray hard band flux for our volume limited sample of active galaxies. The higher the value of this ratio, the less affected our measure of stellar mass will be by contamination of light from the AGN. The mean value for galaxies in our sample is 3.4 (median 2.9), implying that less than a 1/3 of the light in the K band will have been contributed from the AGN on average across our sample (assuming a globally flat spectrum). Therefore, the majority of the light emitted will be from the host galaxy within this frequency range. Nevertheless, there remains a fraction of $\sim$ 15\% of AGN with  $\nu f_{\nu}$ (K band) / $\nu f_{\nu}$ (2-10KeV) $<$ 1.5 where significant infiltration of AGN light on the galaxy SED is possible. The implications of which are discussed in \S 3.1.}
\end{figure}

Moreover, we examine the SEDs for our sources in other ways, specifically by taking the ratio of the K-band luminosity to that of the hard band X-ray luminosity to determine the possible extent of AGN contribution to the optical SED. The K-band represents, roughly, rest frame R band at z $\sim$ 1 and rest frame V band at z $\sim$ 2. In general one would expect the light to be dominated by the underlying stellar population (rather than AGN) at these wavebands (see e.g. Mushotzky et al. 2008). Furthermore Mushotzy et al. (2008) and McKernan (2010) analyse the correlation between infra red light and X-ray hard band luminosity for samples of low redshift AGN, concluding that there is a very steep dependence (suggesting AGN contamination on the galaxy SED) but only for infrared fluxes taken from the very centre of the galaxies in question. Since we disregard point sources, selecting only extended galaxy-like objects, we reduce this possible systematic somewhat. This not withstanding, we run our own analyses of the ratio between K-band luminosity and X-ray hard band luminosity to be assured of the reliability of our stellar mass estimates. If this ratio is very low, it might imply that the source has an SED which is dominated by the AGN, leading to an overestimate of the stellar mass of the host galaxy.

We find that the mean $\nu f_{\nu}$ (K band) / $\nu f_{\nu}$ (2-10KeV) ratio is 3.4 (with a median value of 2.9) for our volume limited sample of `Seyfert' galaxies, suggesting that the majority of light in the K band is from stellar light, not AGN. For example, for QSOs one would expect this ratio to be less than 1.5 - 2 (see Green et al. 2009). The percentage of objects with $\nu f_{\nu}$ (K band) / $\nu f_{\nu}$ (2-10KeV) $<$ 1.5 is less than 15\%, and for all other objects the light in the K-band must be dominated by stellar light, minimising any contributive uncertainty from the AGN in the stellar mass determination. A frequency histogram of the $\nu f_{\nu}$ (K band) / $\nu f_{\nu}$ (2-10KeV) ratio is presented in Fig. 3. As mentioned before, and as described in Bundy et al. (2008), we do not find within the stellar mass fits a population of galaxies which are not well fit by the stellar population synthesis models which we use. As described in Bundy et al. (2008), when examining the minimum $\chi^{2}$ for these fits, the galaxies with AGN have a slightly higher value than those which are not, yet the fraction with $\chi^{2} > 10$ is only 10\%. Furthermore, there is no correlation between the $\chi^{2}$ value and the luminosity of the X-ray AGN (Bundy et al. 2008), a further indication that the light from these galaxies is not dominated by the non-thermal AGN component. 

Taken in aggregate, the fact that we discard point sources, select objects to be galaxy like in terms of colour, fit SEDs with low $\chi^{2}$ values to all our galaxies, observe a mean $\nu f_{\nu}$ (K band) / $\nu f_{\nu}$ (2-10KeV) ratio of 3.4 (implying that less than 1/3 of the K band light can be from the AGN on average, assuming a globally flat spectrum), and have good agreement between our photometric redshifts and the spectroscopically confirmed redshifts in our sample (see Bundy et al. 2008, and \S 3.2) gives us high confidence in calculating stellar masses for our sample of active galaxies, at z $<$ 1.5 at least.

At z $>$ 1.5 we still select galaxies to be extended sources with galaxy like colours, and note that all are fit well by star formation history SEDs with low $\chi^{2}$ to the data. Nevertheless, we have no spectroscopically confirmed redshifts at z $>$ 1.5 within our X-ray selected volume limited sample, so we cannot compare these with our photometric redshifts directly. However, previous studies comparing spectroscopic and photometric redshifts at high z find that there is a probable systematic offset whereby photometric redshifts underestimate the true spectroscopic redshifts (see Bundy et al. 2008, Aird et al. 2009, and \S 3.2 for a thorough discussion). Therefore, the stellar masses in our highest redshift bin must be taken with more caution, with X-ray luminosities systematically lowered due to this effect. 

However, we find that the mean hard band X-ray luminosity of our volume limited sample of massive galaxies rises with redshift, an effect in opposition to the systematic lowering of photometric redshifts for high redshift AGN host galaxies (discussed in more detail in \S 3.2 below), which leads to general confidence in the trend of our results despite this possible bias. Furthermore, our computed X-ray luminosities will be minima for this regime (due to the systematic underestimation of the redshifts of the sources), and in the later discussions in this paper we consider them as such, computing further minimum and maximum quantities, such as the mean maximum lifetime of AGN in our sample. We turn to sources of systematic error on photometric redshifts and hence stellar masses and X-ray luminosities in the next section.

From our discussion above it is evident that there is, however, a population of galaxies for which AGN contamination in the SED of the galaxy could lead to a systematic overestimation of the galaxy's stellar mass. This effect is in opposition to the systematic underestimation of stellar masses and X-ray luminosities for AGN at high redshifts, due to underestimating photometric redshifts (discussed briefly above, and in more detail in \S 3.2 below). To model this first effect, we restrict our sample to those active galaxies with $\nu f_{\nu}$ (K band) / $\nu f_{\nu}$ (2-10KeV) $>$ 1.5 and note that this leads to no significant deviation in the mean stellar masses of host galaxies calculated (to within 1 $\sigma$) and thus no significant departure for the latter derived quantities under discussion in this paper. Thus, the dominant bias in our results must be in underestimating stellar masses and X-ray luminosities due to our photometric redshift estimates being systematically lowered for AGN at high redshifts. It is to this effect we turn to in the next section.

\subsection{Photometric Redshifts of X-ray Sources: Systematics}

One of the major issues with measuring photometric redshifts for X-ray sources is that these X-ray sources could have SEDs
which are contaminated by the AGN, producing inaccurate photometric redshifts. For our lower redshift data in the EGS field this has been extensively investigated by Bundy et al. (2008), who find a $\delta$z = dz/(1+z) $\sim$ 0.11 for the whole sample of X-ray selected galaxies, and no evidence of additional systematic offsets out to z $\sim$ 1. At higher redshifts than this, Bundy et al. (2008) note that there is a systematic offset whereby photometric redshifts underestimate the true spectroscopic redshifts for X-ray luminous AGN. This effect is most prominent for QSO luminosity sources with $L_{X} > 10^{44}$ erg s$^{-1}$, which are predominantly excluded from our sample due to selecting against point sources, and objects without galaxy like colours. In fact less than 10 \% of our sample of AGN are in this luminosity regime and none are point-like. However, as seen by Bundy et al. (2008), there is still an observed modest systematic offset for `Seyfert' type AGN with $L_{X} > 10^{43}$ erg s$^{-1}$ at z $>$ 1 in the EGS field. In fact $\sim$ 15 \% of these sources at 1 $<$ z $<$ 1.5 lie outside the spectroscopic - photometric redshift dispersion of $\delta$z $\sim$ 0.11.

For our high redshift data in the GOODS-N/S fields at 2 $<$ z $<$ 3 the effects of this systematic trend in photometric redshifts underestimating the actual spectroscopic redshifts for AGN is potentially much more significant. Aird et al. (2009) investigate the reliability of photometric redshifts for AGN in the GOODS fields finding that there is accord between spectroscopic and photometric redshifts out to z $\sim$ 1.2 for their sample, in agreement with Bundy et al. (2008). At redshifts higher than this they describe a systematic tendency for the photometric redshifts to be too low in value, which leads to a catastrophic failure in the photometric redshift codes for these objects. None the less, Aird et al. (2009) also note that the worst affected photometric redshifts are for QSO sources with $L_{X} > 10^{44}$ erg s$^{-1}$, of which we only have a few in our sample. With these X-ray luminous sources removed, they find a $\delta$z = dz/(1+z) $\sim$ 0.13 for the remaining galaxies. However, there remains a small systematic bias, leading to an underestimation of the redshifts of lower luminosity AGN at z = 2 - 3.

We investigate this issue with Monte-Carlo simulations of how uncertainties in photometric redshifts lead to uncertainties in our results. In our random (Gaussian) error analyses we assume a $\delta$z = dz/(1+z) $\sim$ 0.2 for all of our redshifts which is higher than that found by both Aird et al. (2009) and Bundy et al.(2008). Since there are no outliers in the spec-z - photo-z plane (with $\delta$z $>$ 0.2) for AGN in our sample at z $<$ 1, we have high confidence in our X-ray luminosities and stellar mass estimates in this redshift range, to within the random error margins intrinsic to calculating stellar masses and luminosities, as plotted in the latter figures of this paper. We ran a standard symmetric random Monte-Carlo simulation to deduce the errors of stellar masses, X-ray luminosities and Eddington masses arising from random errors of $\delta$z = 0.2 on the photometric redshifts. These are incorporated into our confidence claims.

In our intermediate (1 $<$ z $<$ 1.5) and higher (1.5 $<$ z $<$ 3) redshift ranges we minimise systematic effects by actively selecting against QSOs, but acknowledge that some bright point sources probably still remain to some level. It is pertinent here to note that these additional errors are in one direction only, i.e. they systematically lower our photometric redshift from its true value. This, therefore, gives us a minimum redshift and, hence, a minimum X-ray luminosity for our higher redshift AGN, leading to a minimum estimate of the Eddington limiting mass. Additionally this effect will systematically lower the stellar masses of the host galaxies measured, leading to us effectively deducing a minimum stellar mass of very high redshift and X-ray luminous host galaxies. Since much of the analyses of this paper concern maxima and minima values of quantities such as the lifetime of AGN, the luminosity density due to AGN, and the fraction of active galaxies, we can proceed adequately with this limitation. Moreover, perhaps one of the most significant trends we observe in this paper is a modest rise in average X-ray luminosity (and, hence, Eddington limiting mass) with redshift. This trend is in the opposite direction to the systematic bias lowering our photometric redshifts, and thus X-ray luminosities, so we retain confidence in this result despite the possible systematics.

This notwithstanding, we ran a further set of Monte-Carlo simulations to investigate possible asymmetric effects of this systematic bias on the measures of stellar masses, X-ray luminosities and the Eddington limiting masses of SMBHs in our sample. As discussed above, no biases are expected for the lowest redshift bin (z $<$ 1), and all random errors quoted can be considered free from systematic additions here. For the two higher redshift ranges we consider systematic deviations to lower redshifts. For the intermediate redshift range we consider the spec-z - photo-z plot in Bundy et al. (2008), noting that $\sim$ 15 \% of active galaxies have their photometric redshifts systematically lower than their actual spectroscopic redshifts. We allowed 15 \% of our sources to experience a random systematic shift to higher redshift up to the limit of the outliers in this redshift range, assuming a Gaussian distribution of outliers between the limit of $\delta$z = 0.2 and the maximum outliers location. From this Monte-Carlo analysis we recompute stellar masses, X-ray luminosities and (minimum) Eddington masses for the SMBHs and host galaxies in our sample finding that there is a systematic bias to lower photo-z's of $\delta$z = 0.09 leading to a systematic lowering of stellar masses of 0.14 dex, subdominant to the random errors (estimated conservatively to be +/- 0.3 dex in Bundy et al. 2008 and Conselice et al. 2010 in prep.).

For our highest redshift range we have no spectroscopically confirmed redshifts within our volume limited sample. Nonetheless, we have analysed those active galaxies within the redshift range with spectroscopic redshifts that lie outside the volume limited sample and find no significant offset between their photometric and spectroscopic redshifts. This was done for a very small sample, however, and we defer to the more comprehensive test of this bias in Aird et al. (2009). Aird et al. (2009) find that at z $>$ 1.2 there is a significant systematic lowering of photometric redshifts compared to spectroscopically confirmed ones. We model this effect by allowing all sources in our volume limited sample at this high redshift range to be randomly distributed between the limit z$_{spec}$ = z$_{phot}$ to the location of the lowest outlier. This leads to a systematic lowering of photo-z's of $\delta$z = 0.28 leading to a systematic shift in stellar mass of $\delta M_{*}$ = - 0.38 dex (downwards). Here the systematic effect of miscalculating photometric redshifts to be lower than their actual values for AGN at high redshifts comes to dominate over the random errors inherent in measuring photometric stellar masses for high redshift galaxies.

In all of the redshift ranges the errors deduced from random and systematic error analysis through Monte-Carlo simulation come to dominate for the X-ray luminosities and Eddington masses, over the intrinsic errors in their fluxes. This is to say that the errors on photometric redshifts are always more significant than those on the fluxes for our sources. The errors in photometric redshifts and stellar masses, deduced from a combination of random and asymmetric Monte-Carlo simulations (described above), lead to errors in the Eddington limiting black hole mass which range from: $+/- 2 \times 10^{6} M_{\odot}$ at $z = 0.4 - 1$, $+3.2 -3 \times 10^{6} M_{\odot}$ at $z = 1 - 1.5$ and $+6 -4 \times 10^{6} M_{\odot}$ at $z = 1.5 - 3$. The Eddington mass is a lower limit to the mass of the SMBH, given its bolometric luminosity. To compute this we use a bolometric luminosity chosen to be a minimum from the literature and our own analyses (see \S 4.2), thus ensuring that our measured Eddington masses are indeed minimum masses.

\subsection{Completeness}

In order to probe possible trends of SMBH evolution with redshift we must be extremely careful to be complete in both X-ray luminosity for the AGN, and stellar mass of the host galaxy, within whatever redshift range and survey area we choose to investigate. Thus, it is essential to obtain an optimal compromise between depth, area, stellar mass and X-ray luminosity threshold. In this study we are complete to $L_{X} > 2.35 \times 10^{43}$ erg s$^{-1}$, and $M_{*} > 10^{10.5} M_{\odot}$, for the reduced area of the EGS and GNS fields we restrict our redshift analysis to. Thus we effectively probe high redshift counterparts to local Universe `Seyfert' luminosity AGN in average massive elliptical galaxies. The higher mass galaxies probed in our sample may also be progenitors for brightest cluster galaxies (BCGs) in the local Universe. The areas to which we acheive completion are approximately 1/2 of the AEGIS-X field (1/4 of the total EGS POWIR field) and 2/3 of the GNS field (1/9 of the total GOODS ACS North and South fields), corresponding to the regions where the Chandra X-ray data is deepest, and the off axis angle of the Chandra pointings are minimised (see Table 1, and Figs. 1 and 2). We also may miss active galaxies due to their accretion discs having torus opening angles oriented so that light must travel through the Compton thick accretion disc to reach us. In part this is reduced by us using exclusively the hard Chandra bands in our analysis where we will be least sensitive to this obscuration, however, it is likely that some active galaxies will be missed due to this effect. In fact, as seen by Ueda et al. (2003) and Steffen et al. (2003), there are relatively few Compton thick souces detected in the local Universe, even when probed with energies up to $\sim$ 200 KeV. Thus, for this and other reasons (including our use of a minimum bolometric correction, see \S 4.2) our measures of the total active fraction of massive galaxies, and our estimates of the energy output due to AGN must be considered as lower limits, as discussed in the later sections of this paper.

In probing the possible evolution in the $M_{BH} - M_{*}$ relationship at higher redshifts we have been careful to achieve an optimal compromise between minimising various biases, and performing a detailed error analysis of those remaining. The lower the luminosity cut we utilise for our volume limited sample the smaller the area that the X-ray surveys are complete to. Additionally, the brighter the X-ray sources, the higher the contamination of optical light will be on the SED of the galaxy from the AGN and the greater the systematic lowering of the photometric redshifts, leading to less robust stellar mass estimates of the host galaxy (c.f. \S 3.1 - \S 3.2). Taking both of these issues seriously, we have found a suitable compromise. We probe a range of X-ray luminosities for which we are complete out to z $\sim$ 3 for a significant fraction of the area of the two main surveys we investigate (EGS and GNS) whilst still being populated by AGN with low (on average $<$ 1/3) stellar mass contamination from the AGN. Additionally, there remain reasonably small biases due to the systematic offset of photometric redshifts, which we model and factor into our error calculations. As such, we select bright (non-QSO) `Seyfert' galaxies with L$_{X}$ $>$ 2.35 $\times$ 10$^{43}$ erg s$^{-1}$ and in practice L$_{X}$ $<$ 5 $\times$ 10$^{44}$ erg s$^{-1}$ for our volume limited sample. Within this order of magnitude range in X-ray luminosities we are complete in X-ray sources out to z $\sim$ 3 and furthermore have strong estimates of host galaxy stellar masses, which will not be dominated by contamination from the AGN, or unreasonably disturbed by the systematics involved in measuring photometric redshifts for high redshift AGN. The remaining random and systematic errors and biases are modelled through Monte-Carlo simulations (see \S 3.2) and factored into our plots and confidence calculations on our claims.

\section {Method}
\subsection{Detecting AGN}

The purpose of this paper is to explore the co-evolution of supermassive black holes and their host galaxies for a statistically significant number of AGN out to very high redshifts (z $\sim$ 3). In order to acheive this we have combined two large galaxy catalogs (see \S 2.1): the POWIR survey in the EGS field which amasses over 20,000 galaxies at intermediate redshifts (0.4 $<$ z $<$ 1.5); and the GNS which images over 8,000 galaxies as part of an ultra-deep `pencil beam' high redshift (1.5 $<$ z $<$ 3) galaxy survey. We combine these galaxy surveys with the deepest available X-ray data from the Chandra X-ray Observatory overlapping these original galaxy survey fields. We select AGN in this study on the basis of high X-ray fluxes. For all detections we use the same two criteria for determining whether or not a given galaxy has an active SMBH at its centre. A given galaxy is deemed `active' if:

\begin{enumerate}
\item {There is a unique Chandra X-ray detection within a 1.5 arcsec radius of the NIR positional centre of the galaxy.}
\item{The hard band X-ray luminosity is $L_{X}$ $>$ 10$^{42}$ erg s$^{-1}$.}
\end{enumerate}

This ensures that it is most likely that the X-ray source is associated with the host galaxy, and that the X-ray activity is most probably the result of accretion onto a central SMBH, as opposed to being due to stellar formation. Typical X-ray luminosities for massive galaxies without AGN vary widely due to differing star formation rates, but even those with very high star formation rates are expected to have $L_{X} < 10^{42}$ erg s$^{-1}$, with avergae non-AGN massive galaxies having $L_{X} <<  10^{42}$ erg s$^{-1}$ (see e.g. Reddy et al. 2005). Galaxies in both the EGS and GNS fields are selected to be galaxy-like in terms of their colours, and to be extended sources. This effectively selects against QSOs where stellar mass estimates would be significantly compromised by the AGN, leaving us with a sample of (sub-QSO) active galaxies, for which we can estimate stellar masses. With this criteria we identify a total of 436 AGN from the EGS field and 72 AGN from the GOODS North and South fields, giving in total 508 AGN with photometric and spectroscopic redshifts in the range 0.4 $<$ z $<$ 6. This is our large sample, and suffers from not being volume limited. Thus, we cannot use this sample directly to probe evolutionary traits. However, this sample can be used to investigate intrinsic AGN properties.

In order to study any possible evolution in SMBH mass or luminosity with redshift it is important to construct a volume limited sample, where selection effects are adequately addressed. Out to z = 3 for the GNS and z = 1.5 for the EGS, we are complete for the entire areas of our surveys (EGS and GNS) down to galaxy stellar masses of M$_{*}$ = 10$^{10.5}$ M$_{\odot}$, and complete down to specific X-ray luminosities of $L_{X}$ $>$ 2.35 $\times$ 10$^{43}$ erg s$^{-1}$ for approximately 1/4 of the EGS area (1/2 of the AEGIS-X area) and 2/3 of the GNS area (1/6 of the total GOODS area). The redshift limits are due to the constraints on the depths of both the NIR galaxy surveys and Chandra X-ray surveys used (see \S 2.1 and \S 2.4 for further details), and our desire to construct a sample where we are 100 \% complete (to at least 5 $\sigma$).

In addition to the criteria for selecting an `active' galaxy (above) we impose the following criteria to select our subset volume limited sample of principally `Seyfert' luminosity AGN:

\begin{enumerate}
\item{The hard band X-ray luminosity is $L_{X}$ $>$ 2.35 $\times$ 10$^{43}$ erg s$^{-1}$.}
\item{The stellar mass of the host galaxy is M$_{*}$ $>$ 10$^{10.5}$ M$_{\odot}$.}
\item{The redshifts of active galaxies are at 0.4 $<$ z $<$ 1.5 for EGS sources and 1.5 $<$ z $<$ 3 for GNS sources.}
\item{The active galaxy lies within the area of the survey to which the above limits on stellar mass and X-ray luminosity lead to 100\% completion (at 5$\sigma$), to z = 1.5 for the EGS and z = 3 for the GNS.}
\end{enumerate}

\noindent We use this subset to investigate the evolution of AGN over cosmic time. Since we discard point sources, and objects without galaxy like colours, we in principle select against QSO's (thus ensuring we have high confidence in our stellar mass determinations, see \S 3.1 and \S 3.2 for further discussion on this point) and are left with X-ray bright sub-QSO AGN (where we acheive source completion) which we henceforth describe as being `Seyferts', due to their comparable X-ray luminosities.

Fig. 4 illustrates the redshift, luminosity and host galaxy mass cuts we have made to construct a volume limited sample from our larger AGN sample. Red points are those discarded due to our redshift limits, blue points are those discarded due to our luminosity threshold, green points are those discarded due to the host galaxy stellar mass cuts, magenta points are discarded for lying outside of our completion area, and black points are those included in our volume limited sample. This leaves 85 `Seyfert' galaxies with which we probe the co-evolution of SMBHs and their host galaxies at z $<$ 3. A table displaying the co-ordinates and other derived quantities (such and stellar mass of host galaxy and Eddington mass of SMBH) for our volume limited sample is provided in Table A1 in \S A.

\subsection{X-ray and Bolometric Luminosities}

In order to calculate the bolometric luminosity we must first compute the specific X-ray luminosity in the waveband used. Throughout these calculations we use the hard band flux, because it is much less affected by accretion disc and galactic absorption and obscuration than the soft band. In fact the X-ray hard band corresponds to a rest frame X-ray band of roughly 4 - 20 KeV at z $\sim$ 1, and 6 - 30 KeV at z $\sim$ 2. Furthermore, star formation makes a larger contribution to the soft band than to the hard band, rendering the latter preferable to use in probing high-z AGN. The specific luminosity of the X-ray source is calculated via standard techniques as:

\begin{equation}
L_{X} = 4 \pi f_{X} D_{L}^{2} \times K_{corr}
\end{equation}

\noindent where $D_{L}$ is the luminosity distance, derived for a flat spacetime $\lambda$CDM cosmology,$f_{X}$ is the flux of the source, and $z$ is the redshift of the AGN. The K correction that we apply assumes an intrinsic X-ray spectrum of the form $f_{\nu} \propto \nu^{-0.7}$, as favoured by the literature (e.g. Alexander et al. 2003). This results in a K correction $\propto$ (1 + z)$^{0.3}$.

The specific X-ray luminosity must then be converted to a total bolometric luminosity by assuming a spectral energy distribution, or conversion factor. Since we are interested in computing minimum black hole mass estimates from the Eddington limit method, we should also be careful to apply a minimum bolometric correction. In the literature (e.g. Elvis et al. 1994, Elvis et al. 2002, Hopkins, Richards \& Hernquist 2007, Vasudevan \& Fabian 2007 and 2009) it is established that there is a broad range of possible bolometric corrections, ranging from $\sim$ 15 to 200 or so appropriate generally for AGN. By selecting the extreme low end of the distribution of bolometric corrections computed across the literature, we effectively choose a minimum bolometric correction, which will give rise to a minimum SMBH mass through the Eddington limit method. As such, we provide a robust lower limit for the masses of a sample of SMBHs by adopting the minimum bolometric correction factor of 15 in all cases.

\subsection{Eddington Accretion}

\setcounter{table}{1}
\begin{table*}
 \caption{Average Properties of the AGN Samples}
 \label{tab2}
 \begin{tabular}{@{}cccccc}
  \hline
  \hline
Sample	  & $N_{Gal}$ & $< z >$ & $<$ Log($M_{*}[M_{\odot}]$) $>$ & $<$ Log($L_{\rm{2-10Kev}} [\rm{erg s}^{-1}$]) $>$ & $<$ Log($M_{Edd} [M_{\odot}]$) $>$ \\
\hline
Vol. Lim. (`Seyftert') & 85 & 1.239 & 11.02 & 43.76 & 6.755 \\
Total (All AGN)	       & 508 & 1.343 & 10.90 & 43.29 & 6.285 \\
\hline
\end{tabular}
\end{table*}

\begin{figure}
\includegraphics[width=0.5\textwidth,height=0.5\textwidth]{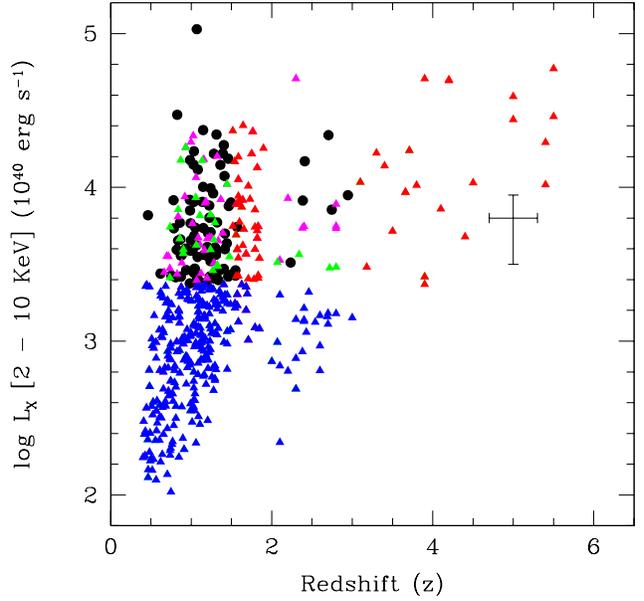}
\caption{Hard band luminosity vs. redshift for the 508 AGN detected in this paper. Red points indicate those AGN excluded from our volume limited sample due to our redshift cuts, blue points are disregarded due to our X-ray Luminosity limits, green points are disregarded for some analyses due to our host galaxy stellar mass cuts, and magenta points are removed due to lying in areas of the X-ray surveys that are not complete to our luminosity threshold. The remaining black points constitute our volume limited sample of 85 AGN with $L_{X}$ $>$ 2.35 $\times$ 10$^{43}$ erg s$^{-1}$, M$_{*}$ $>$ 10$^{10.5}$ M$_{\odot}$ at 0.4 $<$ z $<$ 1.5 \& 2 $<$ z $<$ 3, within the most sensitive regions of the Chandra surveys.}
\end{figure}

In order to determine minimum SMBH mass estimates from our minimum bolometric luminosities we adopt an Eddington limiting method. We follow an approach outlined in detail, for example, in Krolik (1999) and references therein. By setting the outward radiative force equal to the inward force of gravity, we obtain a theoretical minimum mass given the assumption of spherical accretion of Hydrogen gas. Specifically we use:

\begin{equation}
 L_{Edd} = \frac{4 \pi c G M_{BH} \mu_{e}}{\sigma_{T}} = 1.51 \times 10^{38} \frac{M_{BH}}{M_{\odot}} {\rm erg} s^{-1}
\end{equation}

\noindent Where $\sigma_{T}$ is the Thompson cross section of an electron and $\mu_{e}$ is the mass per unit electron. By substituting the observed bolometric luminosity into this above equation and rearranging we obtain an expression for the minimum Eddington mass (M$_{E}$):

\begin{equation}
\frac{M_{E}}{M_{\odot}} = \frac{L_{Bol} ({\rm erg s^{-1}})}{1.51 \times 10^{38}} .
\end{equation}

\noindent The actual mass of the SMBH will be related to the Eddington minimum mass via a simple relation:

\begin{equation}
M_{BH} = M_{E} / \mu
\end{equation}

\noindent where

\begin{equation}
\mu = L_{Bol} / L_{Edd}
\end{equation}

\noindent This is such that the Eddington mass (which can be computed directly from the bolometric luminosity) is equal to $\mu M_{BH}$, where $\mu$ is effectively the efficiency of the SMBHs accretion: the fraction of the bolometric luminosity to the Eddington maximum. If $\mu$ = 1, the SMBH is at the limit at which it can hold onto its accretion disc. If, for example, $\mu$ = 0.1, this implies that the SMBH is radiating at 10\% of its Eddington limit. In the local Universe the average value of $\mu$ is found to be between 0.01 and 0.05 (e.g. Marconi et al. 2004).

The mass accretion rate of a SMBH may be calculated by:

\begin{equation}
\dot{M} = \frac{dM_{BH}}{dt} = \frac{L_{Bol}}{c^2 \eta}
\end{equation}

\noindent Where $\eta$ is the efficiency of mass transfer into electromagnetic radiation. If $\eta$ = 1, there is 100\% transfer of mass into radiation. The parameter $\eta$ is restricted in value via theoretical arguments, and may in fact vary from 0.07 - 0.36 (see Thorne 1974), with an expected mean of $\eta$ = 0.1 which is not predicted to evolve with redshift (Elvis et al. 2002). This has been observationally confirmed in a variety of studies, including Yu \& Tremaine (2002), Elvis et al. (2002) and Marconi et al. (2004).

In this paper we factor $\mu$ out of our mass estimates, plotting $\mu M_{BH}$ (= $M_{E}$) instead, and display the local relation as a function of $\mu$. We take $\eta$ = 0.1 at all redshifts in line with the latest local Universe observations and theoretical arguments (Thorne 1974, Elvis et al. 2002), but note that this may provide an additional source of uncertainty in our results if $\eta$ is permitted to evolve with redshift. We are now in a position to derive lower bounds on the SMBH masses ($M_{E} = \mu M_{BH}$) and actual mass accretion rates ($\dot{M}$) for the 508 AGN we have detected, and utilise the volume limited sample of 85 `Seyfert' galaxies to probe possible evolution over cosmic time. In the following section we investigate how these properties vary as a function of stellar mass of host galaxy, and as a function of redshift. A table displaying the Eddington masses, X-ray hard band luminosities, stellar mass of host galaxies, and redshifts of our volume limited sample is provided in Table A1 in \S A. The average properties of both the full AGN sample, and volume limited `seyfert' sample are given in Table 2.

\section{Results}

\subsection{The Active Fraction}

\begin{figure}
\includegraphics[width=0.5\textwidth,height=0.5\textwidth]{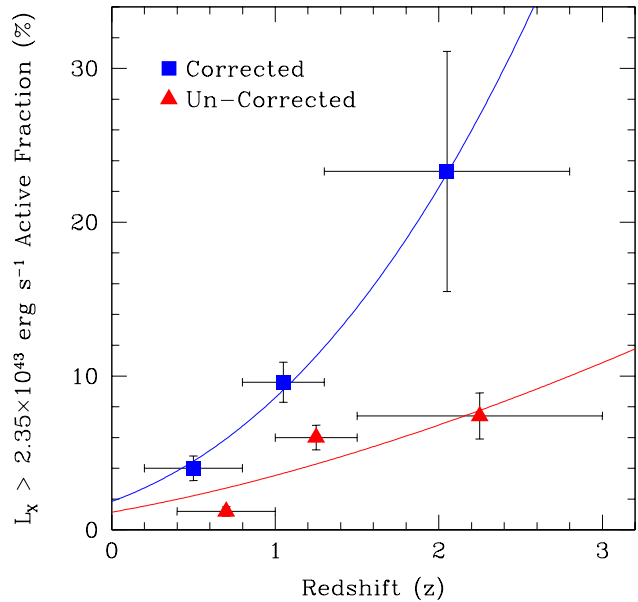}
\caption{The active fraction evolution with redshift. Red triangles indicate active fractions calculated directly from our galaxy sample (section 5.1), with blue squares being minimum corrected active fractions taking into account the maximum amount of time any given galaxy will remain active ($L_{X}$ $>$ 2.35 $\times$ 10$^{43}$ erg s$^{-1}$) and the time interval within the redshift range probed (see section 5.4.2 for more details). The blue squares are shifted slightly to the left for clarity. The solid lines indicate a best fit power law to the data, with exponents of 1.6 +/- 0.3 for the explicit (un-corrected) active fraction evolution, and 2.5 +/- 0.2 for the corrected active fraction evolution. }
\end{figure}

The fraction of a given population of galaxies that are X-ray active above a certain luminosity threshold can prove an interesting probe of both galactic evolution and SMBH formation. In this section we define `active' to be X-ray activity with hard band X-ray luminosities $L_{X}$ $>$ 2.35 $\times$ 10$^{43}$ erg s$^{-1}$, where we are complete to z = 3. We derive the active fraction of massive galaxies by simply computing the fraction of galaxies within the area, redshift, and mass range of our volume limited sample, from the NIR surveys, that have X-ray luminosities above this threshold, as seen by matching to the Chandra surveys. We note here that this will give a sub-QSO active fraction, since point like sources and sources without galaxy like colours are removed from our galaxy catalogs prior to matching, allowing us to compute reliable stellar masses and photometric redshifts (see \S 2 and \S 4.1 for further details). The fractions we compute must be considered as lower limits due to the possibility of missing active galaxies with torus opening angles that lead to very high obscuration even in the hard band probed.

At 0.4 $<$ z $<$ 1.0, for galaxies with $M_{*} > 10^{10.5}M_{\odot}$, we find 23 `Seyfert' galaxies (with $L_{X}$ $>$ 2.35 $\times$ 10$^{43}$ erg s$^{-1}$) in our sample. This is out of a total of 1939 detected galaxies in this stellar mass and redshift range, and within the area covered to completion by both surveys (see Fig. 1 for clarification). Thus, we find 1.2 +/- 0.2 \% of these galaxies to have an active galactic nucleus with $L_{X}$ $>$ 2.35 $\times$ 10$^{43}$ erg s$^{-1}$. This is found to evolve with redshift, such that at 1.0 $<$ z $<$ 1.5 within the same luminosity and mass range we find a total of 53 Seyfert galaxies out of a total number of 889 detected galaxies in this redshift and stellar mass range, and within the area covered to completion by both surveys. Thus, we find 6.0 +/- 0.8 \% of these galaxies to be active with $L_{X}$ $>$ 2.35 $\times$ 10$^{43}$ erg s$^{-1}$ at z $\sim$ 1.25.

For the higher redshift points we use a mix of the Chandra Deep Field North/ South and GNS, where the Chandra data covers the totality of the area probed. At 2.5 $<$ z $<$ 3, for $M_{*} > 10^{10.5}M_{\odot}$, we find a total of 9 `active' galaxies, as defined above. This is out of a total population of 121 galaxies in this redshift and stellar mass range, and within the area covered to completion in both surveys (see Fig. 2 for clarification). Therefore, we find an active fraction of 7.4 +/- 2.0 \% in our high redshift sample. Thus, there is evidence of evolution, whereby the fraction of AGN above our luminosity threshold is larger at higher redshifts in our sample. In fact the active ($L_{X}$ $>$ 2.35 $\times$ 10$^{43}$ erg s$^{-1}$) fraction rises with a 3 $\sigma$ significance between the first two redshift bins, with no significant observed evolution (within the errors) thereafter.

These fractions are considerably lower than the ones deduced in Yamada et al. (2009) for high mass galaxies (M$_{*}$ $>$ 10$^{11}$ M$_{\odot}$) for lower X-ray luminosities. They find 33 \% of their high mass galaxies to have specific X-ray luminosities within the range L$_{X}$ = 10$^{42}$ - 10$^{44}$ erg s$^{-1}$. It is pertinent to note, however, when comparing the results in Yamada et al. (2009) on the active fraction of massive galaxies to the active fractions presented in this paper, that we probe a much higher luminosity regime with $L_{X}$ $>$ 2.35 $\times$ 10$^{43}$ erg s$^{-1}$ used as our selection, where we are far more complete in X-ray sources out to high redshifts. However, our results agree with the active fraction of comparable mass ($M_{*} > 10^{11}M_{\odot}$) galaxies at z $\sim$ 1 of 5 \% calculated in Conselice et al. (2007) using similar X-ray luminosity limits to this paper.

In Fig. 5 we plot the observed active fraction evolution with redshift, and add to this the true (corrected) active fraction evolution (calculated in \S 5.4.2). We note that both the active fraction and corrected active fraction rise with redshift across all redshift ranges probed (except possibly in the highest uncorrected redshift bin). We find a best fit simple power law to the active fraction evolution with redshift of f(L$_{X}$ $>$ 2.35 $\times$ 10$^{43}$ erg s$^{-1}$) = (1.2+/-0.3)(1 + z)$^{(1.6+/-0.3)}$. But note here that it is not a particularly good fit to the data, partially as there is an apparent levelling off at high redshifts.

\subsection{SMBH Mass Evolution}
\subsubsection{The Local $M_{BH} - M_{*}$ Relation}

As described in the introduction, relations have been found to exist between the global properties of galaxies and the SMBHs that reside within their cores. In particular, Kormendy \& Richstone (1995) find a relation between the luminosity of host galaxies and the mass of the SMBHs that reside within them. Furthermore, relations have also been demonstrated to exist between the stellar mass of host galaxies (or bulge mass for disc galaxies) and the mass of their central SMBHs. The most recent and accurate of these relations is found in Haring \& Rix (2004) to be:

\begin{equation}
\begin{split}
\rm{log} (M_{BH} / M_{\odot}) = (8.20 +/- 0.10) \\
+ (1.12 +/- 0.06) \times {\rm log} (M_{*} / 10^{11} M_{\odot})
\end{split}
\end{equation}

\noindent This approximates closely to $M_{*} / M_{BH}$ $\sim$ 1000 for local (z $<$ 0.1) Universe AGN, which we use frequently throughout this paper as an approximation (see also Haring \& Rix 2004).

The purpose of this section is to test whether or not this local $M_{BH} - M_{*}$ relation holds at higher redshifts. To test this we have computed Eddington limiting masses for our volume limited sample of 85 `Seyferts' at z $<$ 3, with reliable stellar mass estimates from optical + NIR data (see Table A1). We plot the Eddington limiting (minimum) mass for the SMBH against stellar mass of the host galaxy in three redshift ranges, 0.4 $<$ z $<$ 1.0, 1.0 $<$ z $<$ 1.5 and 1.5 $<$ z $<$ 3.0 (see Fig. 6). The error bars included represent average composite 1 $\sigma$ errors on the data, derived from Monte-Carlo simulations (see \S 3.2 for details). For the mass points we make no assumption about the value of the Eddington ratio, $\mu$, but plot as lines where the local $M_{BH} - M_{*}$ relation would lie with varying values of $\mu$ from 0.01 to 1.

\subsubsection{Evolution: Assuming a Constant $M_{BH} - M_{*}$ with redshift}

In this section we follow an approach similar to that of Babic et al. (2007) who investigate the Eddington ratios of SMBH accretion discs through assuming a constant $M_{BH} - M_{*}$ relationship with redshift, but expand this from z $<$ 1 to z $<$ 3, increasing the number of sources studied.  Furthermore, by linking AGN to their host galaxies, we also compute the maximum allowable departure from the local $M_{BH} - M_{*}$ relationship at high redshifts. Within this subsection we restrict our analyses to the volume limited sample of `Seyfert' galaxies where we are complete to z = 3, and redshift selection biases are adequately removed.

In order to interpret our $\mu M_{BH}$ vs $M_{*}$ points in Fig. 6 we must make assumptions about either the Eddington fraction, $\mu$, or evolution in the $M_{BH} - M_{*}$ relation. As such, mean values of $\mu$ are calculated across our three redshift ranges, from low to high z, after assuming that the local $M_{BH} - M_{*}$ relation is the same at all redshifts, giving: $\mu$ = 0.056 +/- 0.01 at z = 0.4 - 1, $\mu$ = 0.087 +/- 0.011 at z = 1 - 1.5, and $\mu$ = 0.081 +/- 0.019 at z = 1.5 - 3. In comparison, Panessa et al. (2006) find that massive galaxies in the local Universe have a mean value of $\mu$ $<$ 0.01 across a wide range of X-ray luminosities (larger than the range probed in this paper), and Vasudevan et al. (2009) find a mean value of $< \mu >$ = 0.034 at comparable X-ray luminosities to our sample. Furthermore, we compute an average value for $\mu$ for local Universe AGN, within our X-ray luminosity and stellar mass of host galaxy ranges based on data from O'Neil et al. (2005), finding $< \mu >$ = 0.03 +/- 0.015.

Thus, from an analysis of the mean alone our results suggest some modest increase in $\mu M_{BH}$ vs $M_{*}$ with redshift. This represents a 3 $\sigma$ significant rise in the mean Eddington ratio with redshift between our first two redshift bins, assuming a constant $M_{BH} - M_{*}$ relation at all redshifts, followed by an apparent levelling off from the intermediate to high redshift bin - where the two highest redshift bins have $< \mu >$ equivalent to within 1 $\sigma$. This may be interpreted as a rise in $\mu$ with redshift, requiring no further evolution in the local $M_{BH} - M_{*}$ relation. If this is so, we are witnessing modest evolution in the Eddington fraction, $\mu$, with redshift from $\sim$ 0.03 in the local Universe to $\sim$ 0.09 at z = 1 - 1.5, representing a rise with significance $\sim$ 3 $\sigma$ between z = 0 and z = 1.5 (from Kolmogorov-Smirnoff, KS, test), with no significant further evolution thereafter out to z = 3.

\subsubsection{Evolution: Assuming Constant $\mu$ with Redshift}

Conversely, we can investigate the maximum allowed evolution in the $M_{BH} - M_{*}$ relation, assuming no evolution in $\mu$, for our population of `Seyfert' galaxies within our volume limited sample. Since $\mu$ is intimately related to the total available fuel for an AGN from cool gas, it is natural to expect it will rise at higher redshifts. Thus, a limit can be placed on evolution in the local relation by setting $\mu$ equal to its measured mean value at z = 0.4 - 1 of 0.056, because at redshifts higher than z = 0.4 - 1, $\mu$ is likely $\geq$ 0.056 due to there being, on average, more available gas as fuel for the SMBHs. Therefore, assuming that there is no evolution in $\mu$ with redshift allows us to compute the maximum possible evolution allowable in $M_{BH} - M_{*}$ from our observed evolution in $\mu M_{BH} - M{*}$. From this we compute the maximum evolution in the local ratio $M_{*} / M_{BH}$ as evolving from 1000 at z = 0 (Haring \& Rix 2004) to 700 +/- 100  at z $\sim$ 1.5, less than a factor of 2. It should be stressed that this is a maximum possible departure from the local Universe $M_{BH} - M_{*}$ relation at 3 $\sigma$ confidence to lower values. Some, or all, of this observed departure from the local $M_{BH} - M_{*}$ relation could in fact be driven by evolution in $\mu$, therefore requiring no further evolution in the local relationship. 

Crucially, evolution such that $M_{*} / M_{BH}$ $>$ 1000 in the early Universe is unrestrained in this analysis. We can, however, place an upper limit on evolution to higher, as well as lower, values of $M_{*} / M_{BH}$ by assuming that $\mu$ = 1 in the high redshift Universe. This leads to a maximum positive evolution of approximately a factor of 11 +/- 1.5 (assuming no super-Eddington accretion), indicating that 700 $<$ $M_{*} / M_{BH}$ $<$ 11000 at $z < 3$. The implication of this is that either SMBHs and their host galaxies grow principally together, or else there is dramatic evolution in $\mu$ and SMBHs thus grow after their host galaxies are assembled.

\begin{center}
\begin{landscape}

\begin{figure}
\vspace{3truecm}
\includegraphics[width=8truecm]{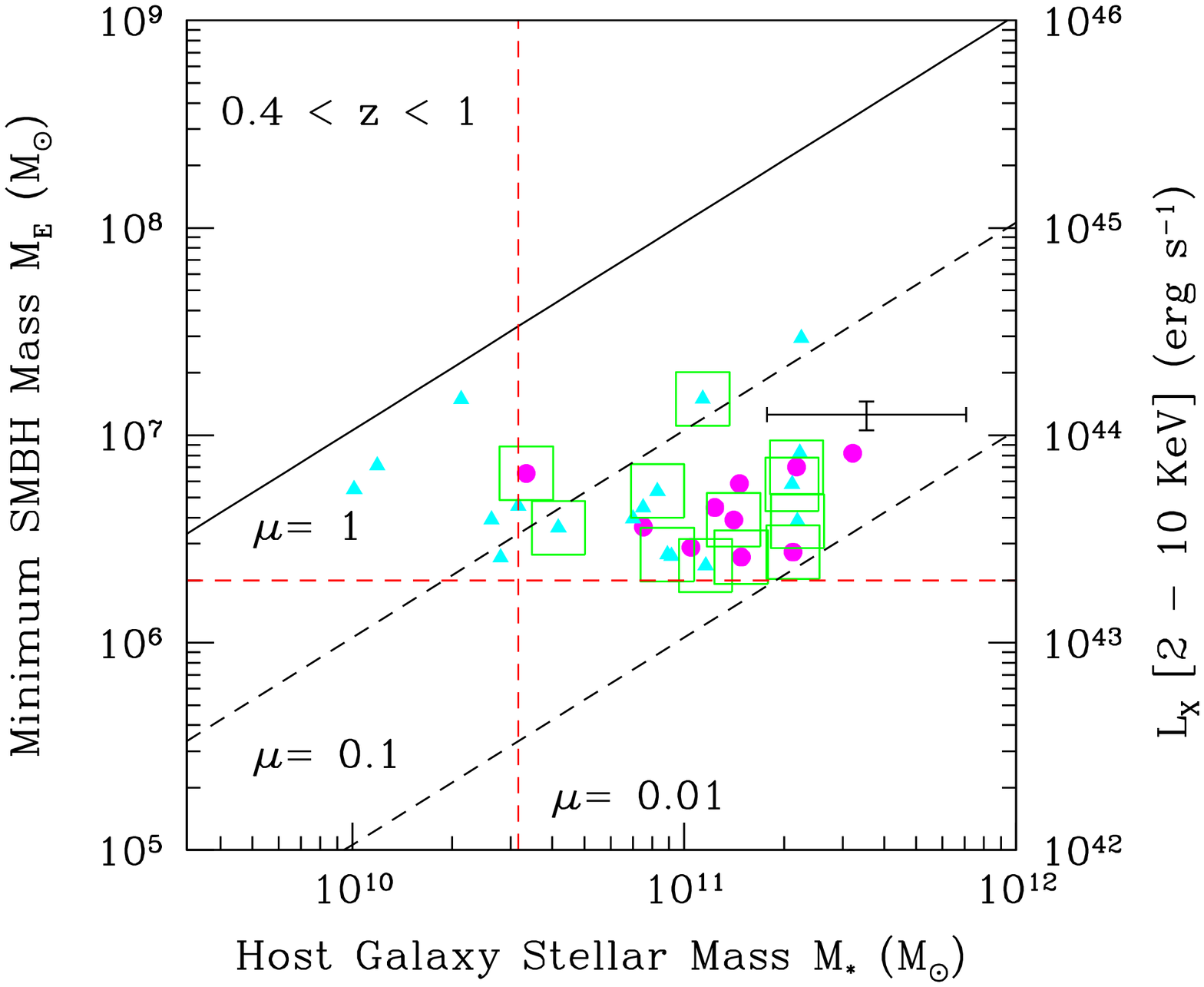} 
\includegraphics[width=8truecm]{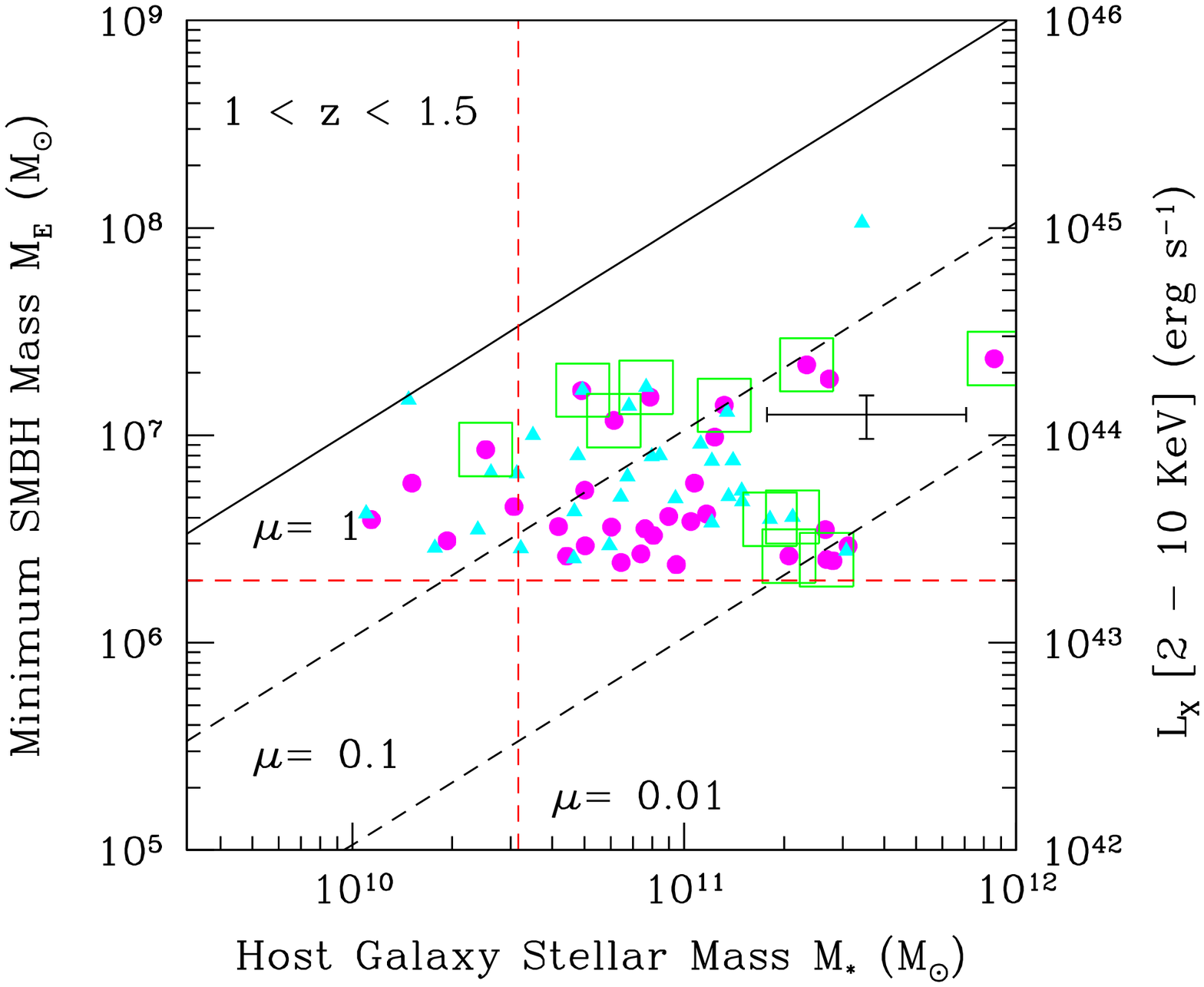}
\includegraphics[width=8truecm]{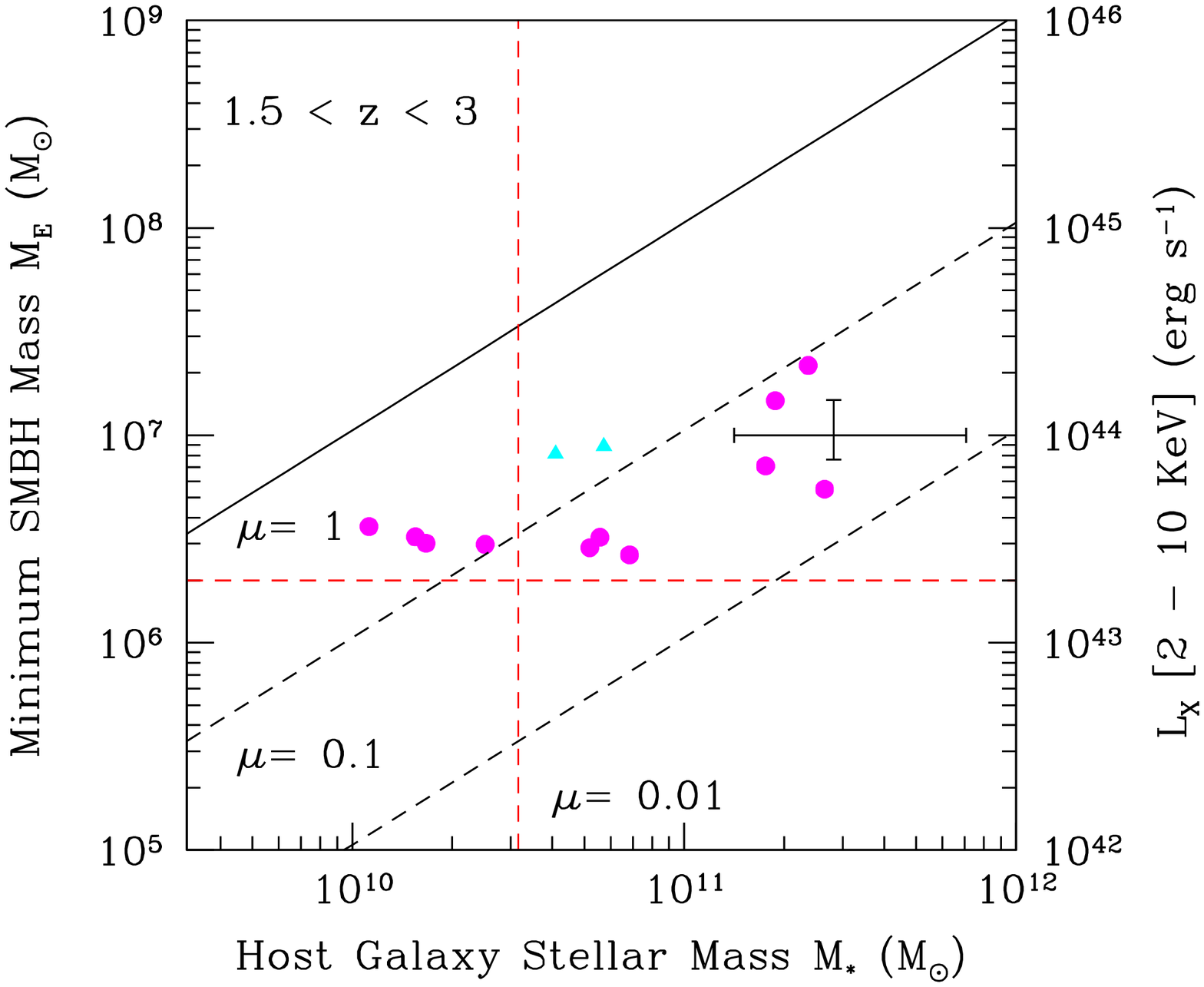}
\caption{The SMBH Eddington limiting (minimum) mass ($M_{E} = \mu M_{BH}$) plotted against stellar mass ($M_{*}$) of the host galaxy across three redshift ranges from left to right, 0.4 $<$ z $<$ 1.0, 1.0 $<$ z $<$ 1.5 and 1.5 $<$ z $<$ 3.0. The error bars represent average 1$\sigma$ errors based on the composite errors in redshift, stellar mass of the host galaxy, and the hard band X-ray flux from Monte-Carlo simulation (see \S 3.2). We also apply a minimum bolometric correction to ensure that our Eddington masses are true minimum black hole masses (see \S 4.2). We plot the X-ray luminosity on the right of each plot for direct comparison to the data. The solid line is the local relation taken from Haring \& Rix (2004), with $\rm{log}(M_{BH}) \sim \rm{log}(M_{*}) - 3$. The dashed lines indicate where the local relation would lie for differing values of $\mu$ in order to compare our data without making prior assumptions about the $\mu$ value. Magenta circles indicate hard X-ray sources with H $>$ 0.5, cyan triangles indicate relatively soft X-ray sources with H $<$ 0.5 (see \S 5.5 for definition). The green boxes indicate objects with spectroscopically confirmed redshifts. The dotted red lines indicate our luminosity and mass cuts for our volume limited sample, where we are complete across all of the redshift bins for the areas probed. Points that lie below the luminosity threshold of our volume limited sample are removed from these plots for clarity. \label{fig. 2}}
\end{figure}

\end{landscape}
\end{center}

\subsubsection{The High Accretion Fraction}

Our mean analysis (above) is somewhat compromised by the fact that at the lower mass end of the host galaxy mass function only AGN with high $\mu$ values are visible above our luminosity threshold. This likely drives the apparent flatness of our $\mu M_{BH}$ vs $M_{*}$ plots compared to the local relation of Haring \& Rix (2004) (plotted in Fig. 6 as a solid line), as galaxies with low stellar masses must have high values of $\mu$ to be visible in the Chandra observations, whereas this restraint is considerably relaxed for the highest mass galaxies in our sample.

To get a better handle on this potential source of bias, we compute the fraction of Seyferts with $\mu$ $>$ 0.1 (f($\mu$ $>$ 0.1)) in each redshift bin, assuming that the local $M_{BH} - M_{*}$ relation is valid at all z. We obtain for ascending z, f($\mu$ $>$ 0.1) = 0.16 +/- 0.03 at z = 0.4 - 1, 0.21 +/- 0.02 at z = 1 - 1.5 and 0.22 +/- 0.07 at z = 1.5 - 3. This indicates a modest 2 $\sigma$ significant rise (from a KS test) in the fraction of galaxies with $\mu$ $>$ 0.1 between z = 0.4 and z = 1.5, from the un-binned data. There is then no significant evolution from z = 1 - 1.5 to z = 1.5 - 3. This potential levelling off may be related to the fact that the total X-ray luminosity from AGN peaks around z $\sim$ 2 for Seyferts in the Universe (see e.g. Barger et al. 2005, Hasinger et al. 2005, Aird et al. 2009), implying that there should be a similar peak in $\mu$ around the same redshift, which is consistent with our results.

In conclusion to this section, our results are consistent with no evolution in the $M_{BH} - M_{*}$ relation with redshift, providing there is mild and plausible evolution in the Eddington ratio, $\mu$, implying that SMBHs and host galaxies grow together throughout cosmic history. Nevertheless, there is a rise in AGN activity represented by an increase in the mean value of the Eddington ratio, $\mu$, out to z $\sim$ 3, and a rise in the active fraction of galaxies (seen in \S 5.1) out to z $\sim$ 3 as well.

\begin{figure*}
\includegraphics[width=0.45\textwidth,height=0.45\textwidth]{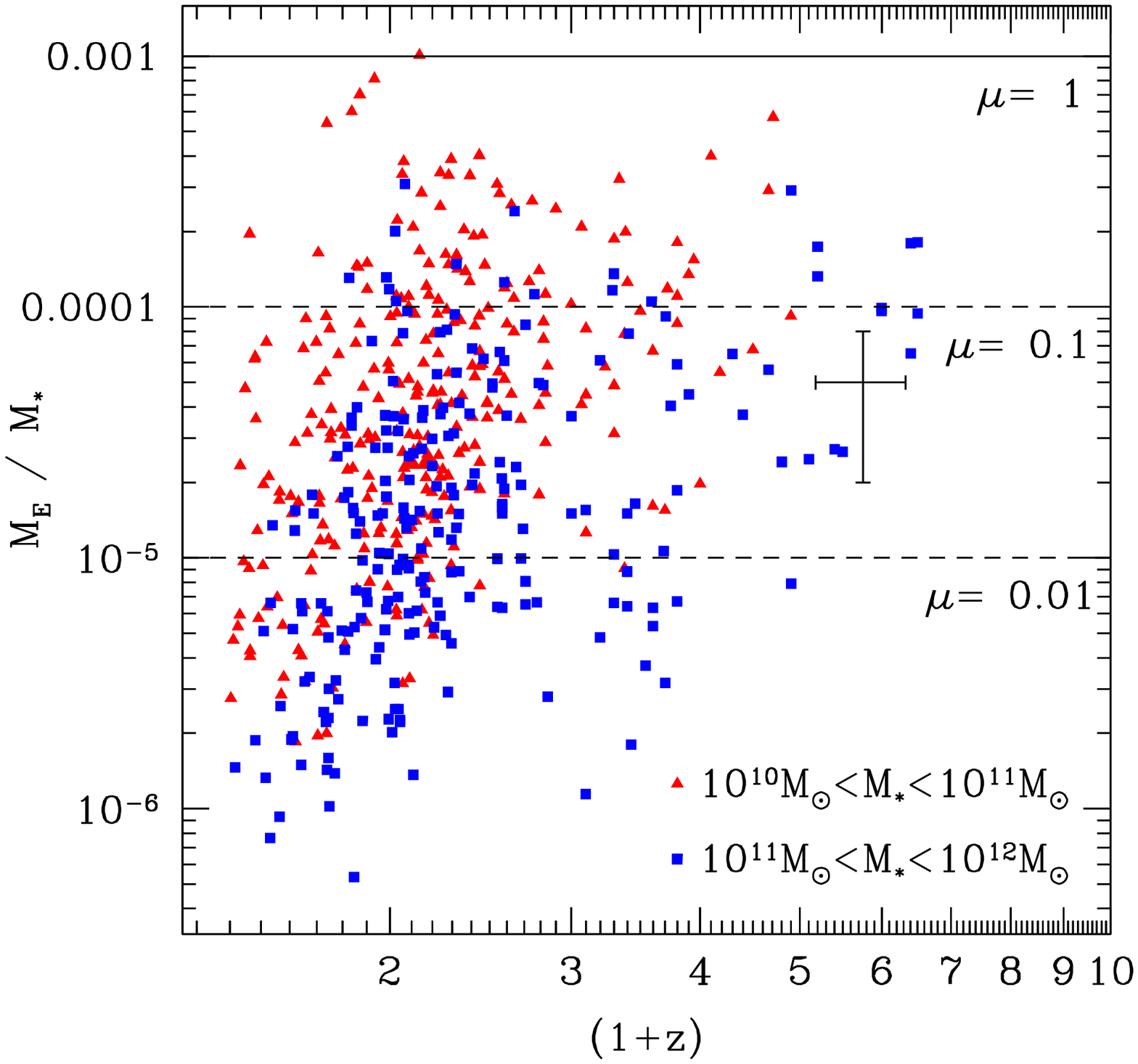}
\includegraphics[width=0.45\textwidth,height=0.45\textwidth]{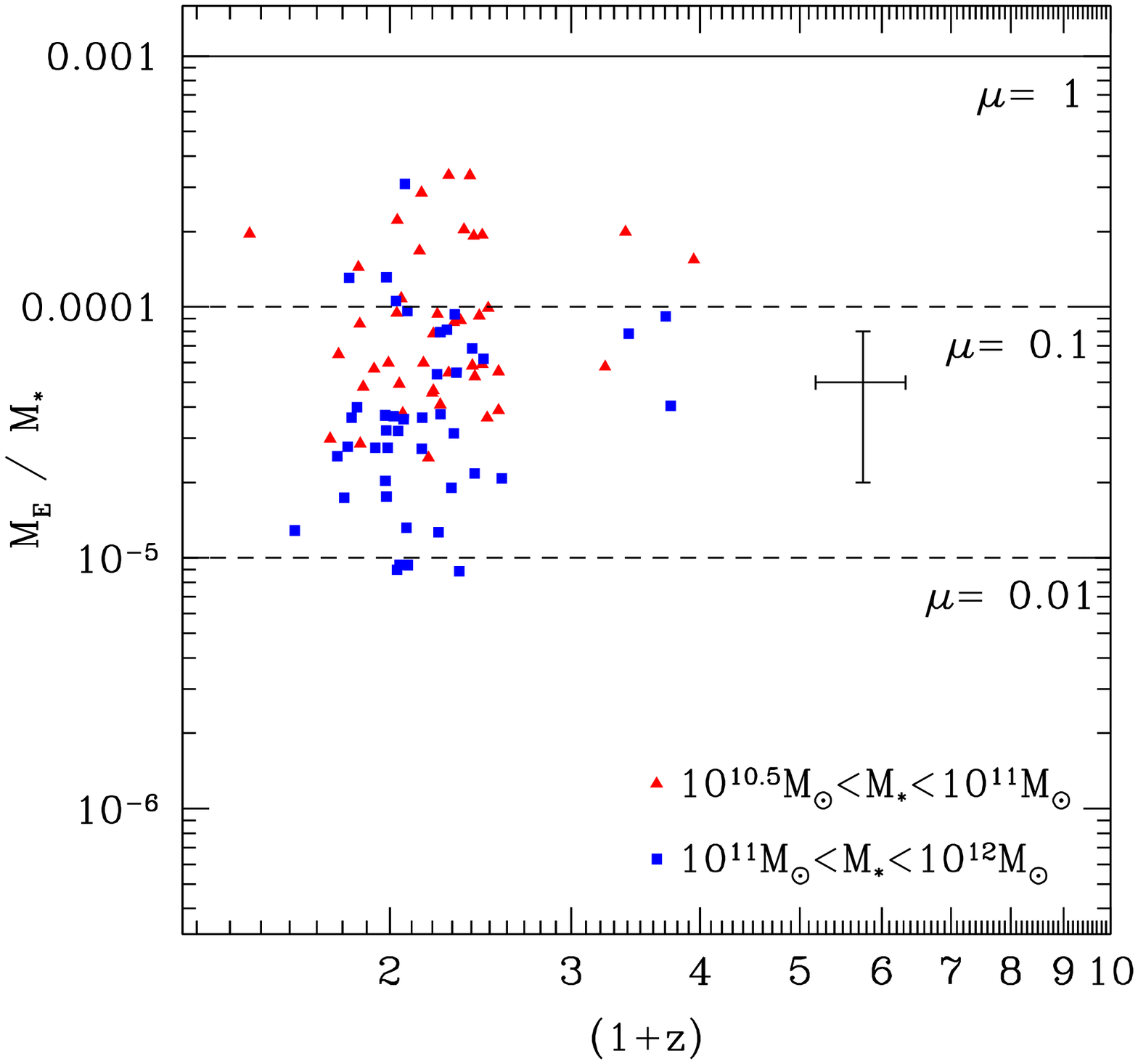}
\caption{The redshift evolution of the ratio of Eddington limiting mass to stellar mass $M_{E} / M_{*}$ = $\mu M_{BH} / M_{*}$. Left plot is for all 508 AGN detected and right plot is for the volume limited sample of 85 `Seyferts'. Red triangles indicate low mass (M$_{*}$ $<$ 10$^{11}$ M$_{\odot}$) host galaxies, with blue squares indicating high mass (M$_{*}$ $>$ 10$^{11}$ M$_{\odot}$) host galaxies. The average 1 $\sigma$ error is plotted in each plot (derived through Monte-Carlo simulation). The solid line represents the local relation of Haring \& Rix (2004), with $M_{BH} / M_{*}$ $\sim$ 0.001. The dashed lines show where the local relation would lie for varying values of $\mu$. Note the selection effect whereby lower mass galaxies have higher $\mu M_{BH} / M_{*}$ ratios at all redshifts. The mean $\mu M_{BH} / M_{*}$ ratio and dispersion of this also varies with redshift in the left hand plot due to selection effects, leading to higher values and lower dispersions at high z. \label{fig. 3}}
\end{figure*}

\subsubsection{Stellar Mass Dependence on $\mu$?}

Fig. 7 shows the evolution of $\mu M_{BH} / M_{*}$ with redshift for the whole sample of 508 detected AGN (left) and the volume limited sample of 85 `Seyferts' (right). On the left hand plot there is an apparent evolution in both the mean value of $\mu M_{BH} / M_{*}$ and the dispersion around the mean. This is largely due to a selection effect driven by the X-ray luminosity limit of the sample. The right hand plot does not show this trend to the same extent because it is volume limited. Nonetheless, there is a tentative rise observed in the mean $\mu M_{BH} / M_{*}$ value in our volume limited sample from our low redshift bin (z = 0.4 - 1) to our intermediate redshift bin (z = 1 - 1.5) at $\sim$ 2 sigma confidence (from KS test), with no significant evolution seen thereafter out to the highest redshift range (z = 1.5 - 3). Both plots, however, show a high degree of scatter around their respective mean values of $\mu M_{BH} / M_{*}$ at all redshifts, which is most probably a result of there being a large diversity in values of the Eddington fraction, $\mu$, in high as well as low redshift AGN (as seen in the local Universe by, e.g., Panessa et al. 2006).

Conversely, if we attempt to fit the volume limited data it becomes clear that our results are consistent with no global evolution in $\mu M_{BH} / M_{*}$, and we can rule out a steep departure from the local relation at high redshifts, provided there is no dramatic evolution in $\mu$ in opposition to the evolution of the $M_{BH} - M_{*}$ relation. This seems plausible because the degree of scatter in  $\mu M_{BH} / M_{*}$ is roughly equivalent across all redshift ranges (when the number of points is taken into account), most probably indicative of there being high variety in $\mu$ values even out to high redshifts.

In both plots, however, there is a clear separation to 3 $\sigma$ confidence (from KS test) between low mass $M_{*} < 10^{11} M_{\odot}$ (red points) and high mass $M_{*} > 10^{11} M_{\odot}$ (blue points) host galaxies. This trend may be largely driven by the fact that low mass host galaxies must have relatively high values of $\mu$ in order to be detected above our luminosity threshold.

To investigate this further we plot $\mu$ vs $M_{*}$ on the left hand plot of Fig. 8, assuming here the validity of the local $M_{BH} - M_{*}$ relation at all z. The values of $\mu$ show a significant decline with increasing stellar mass of host galaxy. The solid line represents the luminosity limit in our sample, and the values of $\mu$ are calculated assuming that the local relation holds at all redshifts. Points below the solid line cannot be seen with our X-ray depth. There does, none the less, appear to be a slight dearth of points with high $\mu$ at the high stellar mass end. This is most likely partially explained as a feature caused by the lack of very massive galaxies in our sample at high redshifts. In this plot we probe the high end of the distribution in $\mu$ and, due to the steepness of the mass function, there are more galaxies populating the low mass end and, hence, it is natural to think that they will reach higher values of $\mu$, simply because there are more of them to randomly populate the plane. Thus, we make no claims regarding possible Eddington ratio, $\mu$, dependence on the stellar mass of host galaxy in this paper.

\subsection{Accretion Rate - Stellar Mass Dependence}

The local $M_{BH} - M_{*}$ relation implies that more massive host galaxies will contain more massive SMBHs in their cores. These may in turn accrete at higher rates corresponding to larger Eddington limiting luminosities compared to lower mass systems. Thus, it is natural to expect that there may be a relation between the accretion rate of SMBHs and the mass of their host galaxies. We investigate this possibility here. Since we witness no strong evolution in the local $M_{BH} - M_{*}$ relation with redshift, with a maximum departure of less than a factor of two to higher SMBH masses for their host galaxy's mass, we investigate the possible dependence of accretion rate on stellar mass across all of the redshift ranges probed.

We calculate the accretion rate for our sample of AGN as described in \S 4.3 (eq. 6). Here we assume a mass to radiation conversion factor of $\eta$ = 0.1 (as in e.g. Alexander et al. 2008), in line with common local Universe estimates, theoretical arguments and global X-ray background measures (see Yu and Tremaine 2002, Elvis et al. 2002 and Marconi et al. 2004). We acknowledge, however, that in this part of the analysis our results are subject to some re-evaluation if the value of $\eta$ is found to vary with redshift and/or mass of the SMBH. We plot these values against their host galaxies' stellar masses in Fig. 9. The faint black dots represent the data, with the average 1 $\sigma$ error on the points from Monte-Carlo simulations also plotted in black. The blue squares represent the mean of the data in each redshift range, with the 2 $\sigma$ error on the mean plotted alongside them. The solid red line indicates the best fit power law function to the data, with dM$_{BH}$/dt = 1.8(+/-0.9) $\times$ 10$^{-11}$ M$_{*}^{0.9+/-0.3}$. There is a high degree of dispersion around the best fit line, and we consequently do not use this fit for any further analysis. The high scatter in both plots is most likely due to varying Eddington ratios ($\mu$) at similar host galaxy stellar mass.

Our principal result here is that there is a high degree of scatter in the $\dot{M} - M_{*}$ plot for all of our detected AGN (left Fig. 9) and for our volume limited sample (right Fig. 9). This not withstanding, there is some evidence from inspection that the most massive host galaxies contain SMBHs with higher accretion rates on average. There appears to be a population of very highly accreting SMBHs present only at the high stellar mass end of the total distribution. This is more or less what one would expect, as in the local relationship one finds higher mass SMBHs in higher mass galaxies and, furthermore, more massive SMBHs may accrete at higher rates before shutting themselves off. Quantitatively, we find a modest increase in mean accretion rate with mass for the whole sample: from 0.0054 +/- 0.0005 M$_{\odot}$ yr$^{-1}$ for M$_{*}$ $<$ 10$^{11}$ M$_{\odot}$ to 0.011 +/- 0.0015 M$_{\odot}$ yr$^{-1}$ for M$_{*}$ $>$ 10$^{11}$ M$_{\odot}$. For the volume limited case we find that the average accretion rate rises from 0.17 +/- 0.02 M$_{\odot}$ yr$^{-1}$ for M$_{*}$ $<$ 10$^{11}$ M$_{\odot}$ to 0.26 +/- 0.06 M$_{\odot}$ yr$^{-1}$ for M$_{*}$ $>$ 10$^{11}$ M$_{\odot}$. From a KS test, the accretion rates of AGN separated via stellar mass, so as to be higher or lower than M$_{*}$ = 10$^{11}$ M$_{\odot}$, form distinct populations to a 3 $\sigma$ confidence level in the total sample, with a weaker, $\sim$ 2 $\sigma$, confidence for the volume limited sample.

\begin{figure*}
\includegraphics[width=0.45\textwidth,height=0.45\textwidth]{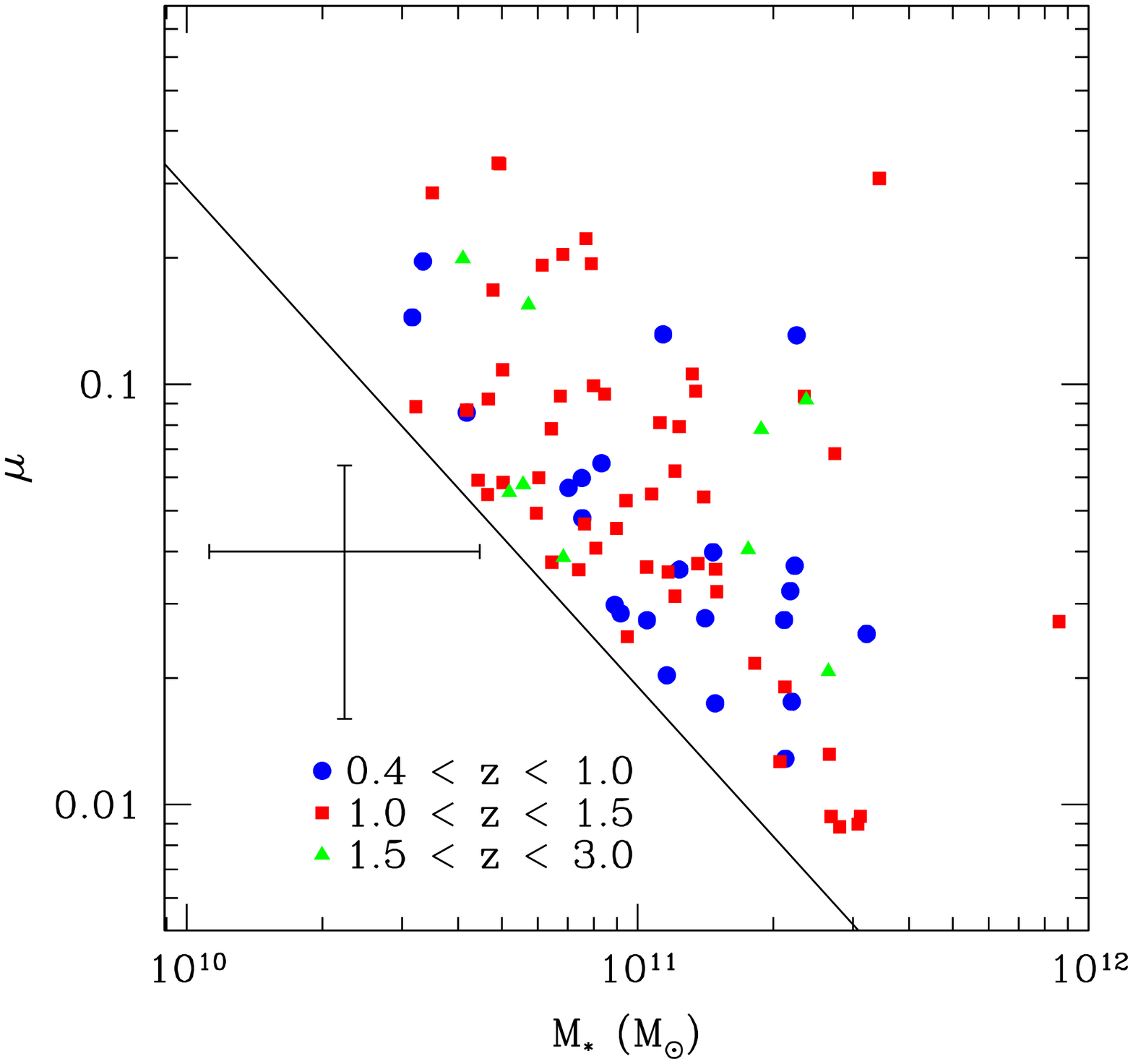}
\includegraphics[width=0.45\textwidth,height=0.45\textwidth]{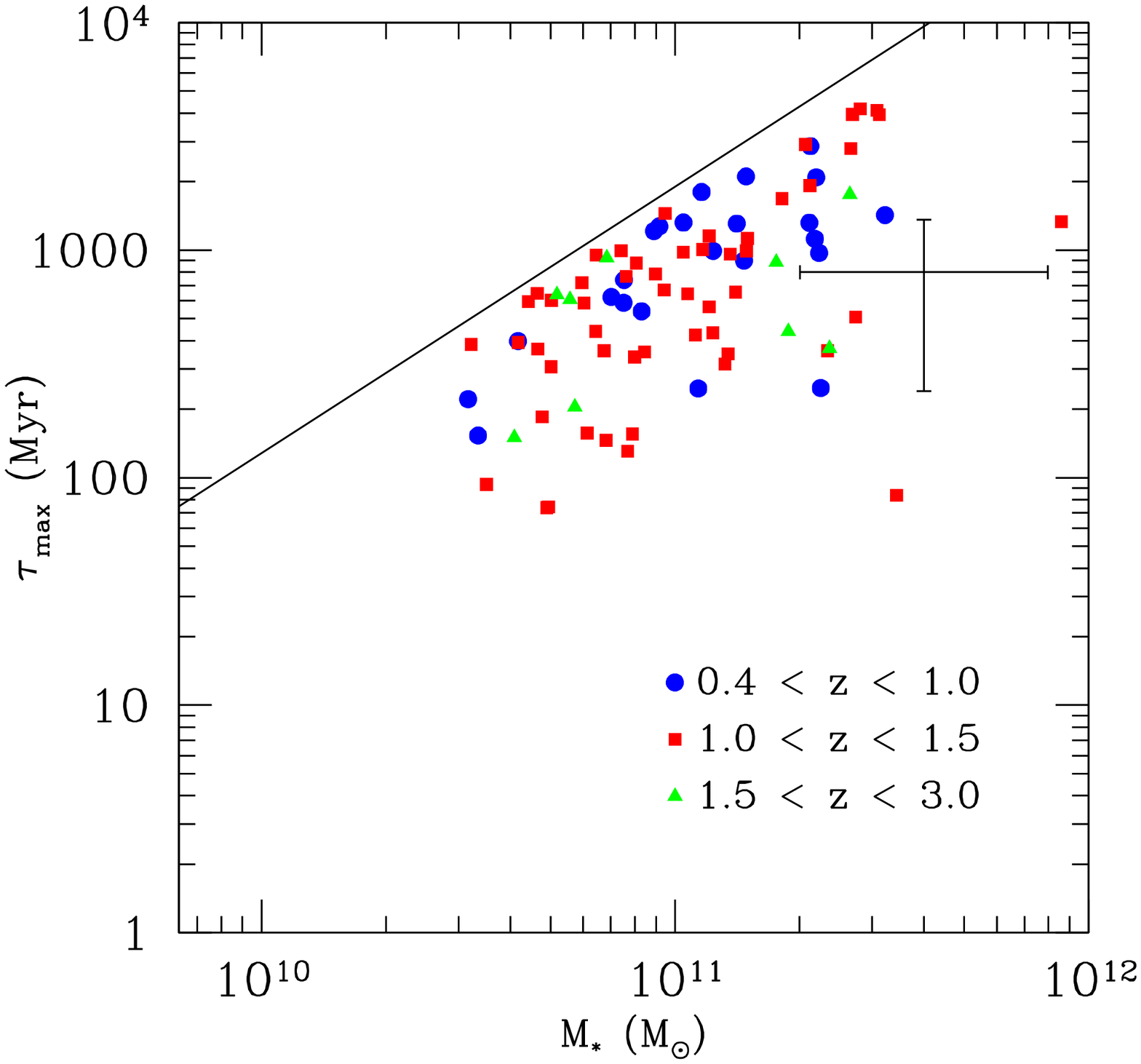}
\caption{Plots of SMBH efficiency $\mu$ (left) and maximum time to reach local relation $\tau_{max}$ (right) as a function of the stellar mass of host galaxy, for our volume limited sample of 85 `Seyfert' galaxies. $\mu$ is calculated assuming the local $M_{BH} - M_{*}$ relation does not evolve with redshift. $\tau_{max}$ is computed by taking $\mu$ = 1, thus assuming Eddington accretion, as lower values of $\mu$ will lead to lower values of $\tau$. The solid line in both plots indicates the luminosity threshold of our volume limited sample. Differing colour and shape points indicate the different redshift ranges being considered, as stated on the plots. In both plots the black error bars represent average 1 $\sigma$ errors on the individual data points.}
\end{figure*}

\begin{figure*}
\includegraphics[width=0.45\textwidth,height=0.45\textwidth]{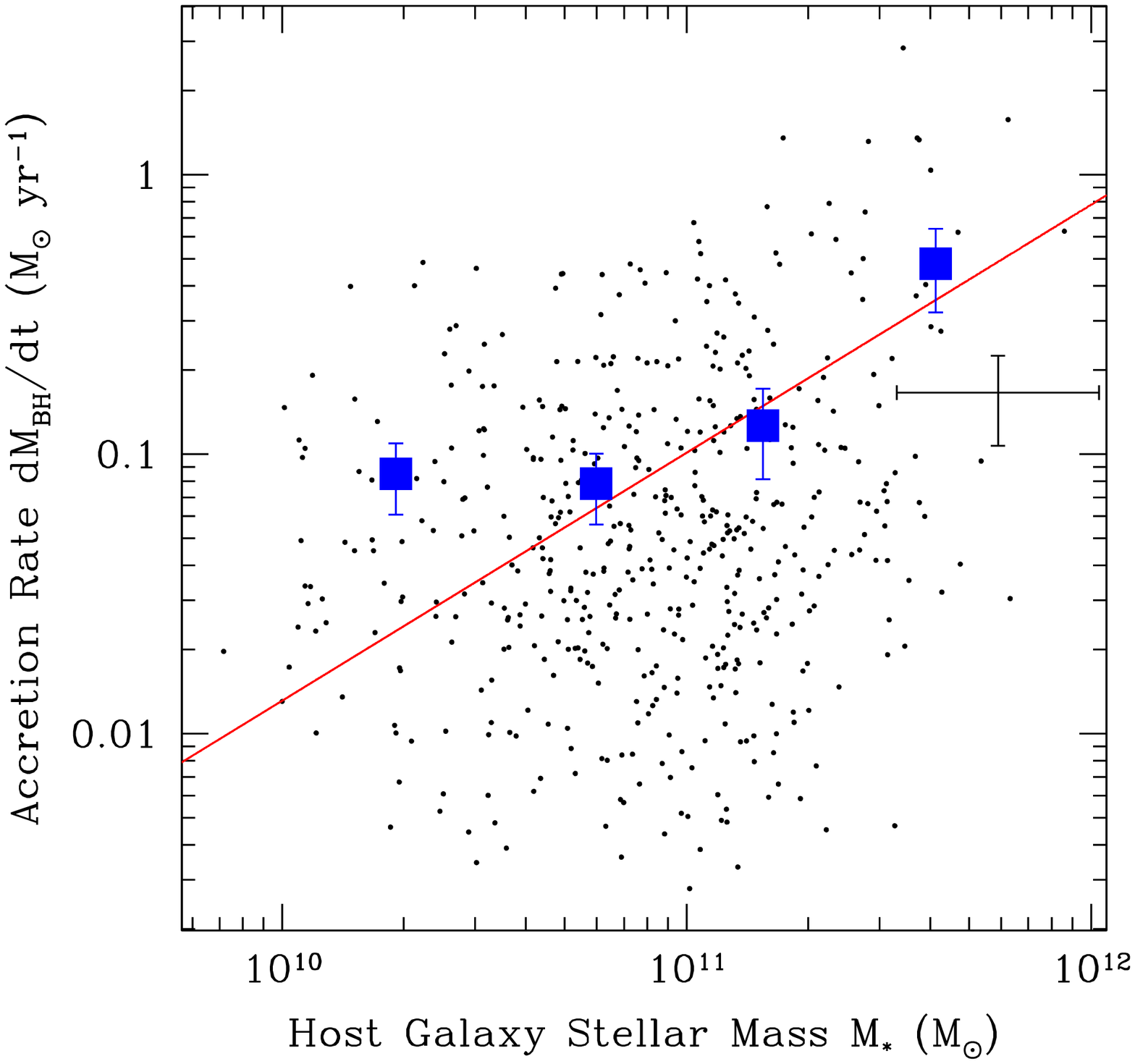}
\includegraphics[width=0.45\textwidth,height=0.45\textwidth]{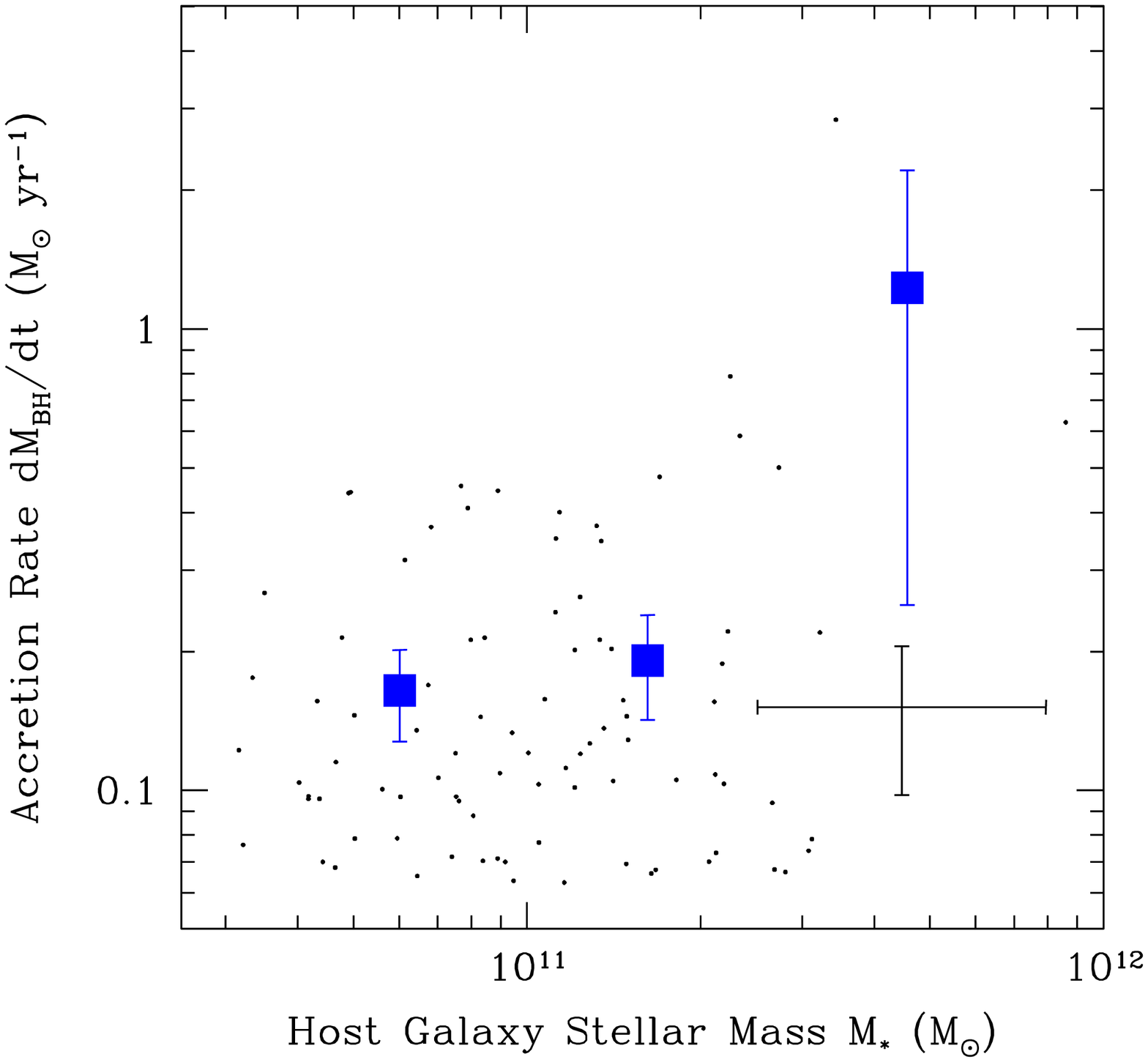}
\caption{On the left is the accretion rate ($dM_{BH}/dt = L_{Bol} / \eta c^{2}$) plotted against stellar mass of host galaxy for all of our detected 508 AGN. We use $\eta$ = 0.1 for all points here to be in line with local Universe values. On the right is the accretion rate ($dM_{BH}/dt$) plotted against stellar mass of host galaxy for our volume limited sample of 85 `Seyfert' galaxies. Blue squares represent the mean value of the accretion rate in each redshift range, with 2 $\sigma$ errors on the mean plotted alongside them. The black error bars in both plots represent the average 1 $\sigma$ errors on the individual data points. The solid red line is a best fit line to the data with dM$_{BH}$/dt = 1.8(+/-0.9) $\times$ 10$^{-11}$ M$_{*}^{0.9+/-0.3}$. Note that very highly accreting AGN are only found at the high stellar mass end in both plots.}
\end{figure*}

\subsection{Global Properties of AGN}

\subsubsection{AGN Lifetimes}

A critical parameter for understanding the evolution of SMBHs is the timescale for which they accrete matter before shutting off. Much work has been performed on constraining the lifetimes of AGN, most commonly for QSOs (e.g. Haehnelt, Natarajan \& Rees 1998, Kauffmann \& Haehnelt 2000, Yu \& Tremaine 2002, McLure \& Dunlop 2004, Martini 2003). These studies find rough accord with theory in predicting that the lifetime of QSOs are in the range 10$^{6}$ - 10$^{9}$ yrs. The large range in possible values for t$_{Q}$ are indicative of a great deal of uncertainty in the field. This is in part because robust estimates of the QSO lifetime, t$_{Q}$, are often dependent on the Eddington fraction, $\mu$, which to determine directly requires independent measures of the SMBH mass to the AGN X-ray luminosity, which frequently proves extremely difficult to achieve for all but the most local active galaxies. 

Measurements of the probable range of lifetimes allowable from the data for lower luminosity AGN, such as Seyfert galaxies, have also been estimated. Constraints on these measures give t$_{S}$ $\sim$ 10$^{7}$ - 10$^{8}$ yrs, perhaps shorter than their brighter QSO counterparts.  In this section we aim to expand this prior work on AGN lifetimes to more modest X-ray luminosities, looking specifically at `Seyfert' luminosity galaxies over a wide range of redshifts. Furthermore, we propose a novel technique for placing an upper limit on the lifetime of AGN by utilising the local $M_{BH} - M_{*}$ relation, and assuming that no galaxy can end up lying above the local relation by z = 0. 

From the accretion rate and measured value of $\mu M_{BH}$ for each SMBH we calculate the maximum time ($\tau_{max}$) that a SMBH can remain accreting at its current rate before it lies above the local relation, by assuming that it is accreting with an Eddington fraction $\mu$ = 1. Values of $\mu$ less than this will lead to shorter timescales as this will imply that the actual SMBH mass is closer to the local limit. Further, we also assume that any increase in mass of the host galaxy (i.e. through major mergers) will result in proportional growth of the central SMBH through corresponding black hole merging. Thus we can write for the maximum time:

\begin{equation}
\tau_{max} = \frac{M_{BH} (z=0) - M_{BH} (z=z')} {\dot{M}} \approx \frac{M_{*}/1000 - M_{E}} {\dot{M}}
\end{equation}

\begin{equation}
= \frac{M_{*}} {1000 \dot{M}} - \frac{\eta \sigma_{T} c} {4 \pi G \mu_{e}} = \frac{M_{*}} {1000 \dot{M}} - 3.75 \times 10^{7} {\rm yrs}
\end{equation}

\noindent Where $\dot{M}$ is the accretion rate of the SMBH, M$_{BH}$ (z=0) $\approx$ M$_{*}$ / 1000 (from the local $M_{BH} - M_{*}$ relation of Haring \& Rix 2004), which is the SMBH's expected mass from the value of its host galaxy's mass at z = 0, M$_{BH}$ (z=z') is the measured minimum black hole mass at each redshift (z') where $\mu$ = 1, M$_{*}$ is the stellar mass of the host galaxy, and M$_{E}$ is the Eddington minimum mass. Since $M_{E}$ and $\dot{M}$ are both proportional to the total bolometric luminosity of the AGN, $L_{Bol}$, the ratio $M_{E} / \dot{M}$ is a constant.  This constant represents the minimum correction to the lifetime of AGN considering the SMBH's current minimum mass and actual accretion rate. Effectively, the ratio $M_{E} / \dot{M}$ is the minimum physically allowable time to construct a black hole of a given mass. It is a constant because higher accretion rates would lead to lower timescales, but higher accretion rates also imply a higher minimum Eddington mass which would require a proportionally longer time to form. The combination of this is that the correction factor is constant.

$\tau_{max}$ indicates the maximum amount of time that any given AGN can maintain its current level of accretion, at its current X-ray luminosity. This quantity is plotted against stellar mass on the right of Fig. 8. One interesting thing to note is that there are some AGN which may continue accreting for a maximum of only a few 100's of Myr at their current rate, with an average maximum lifetime of $\sim$ 0.9 +/- 0.1 Gyr in each redshift range. This value of $\tau_{max}$ implies that t$_{AGN}$ $<$ 10$^{9}$ yr for most systems, which agrees favourably with previous independent measures (using differing techniques), and theoretical expectation (see e.g. McLure \& Dunlop 2004 and Martini 2003 for a comprehensive review of previous work in the field). Moreover, we can use information about $\tau_{max}$ to determine the minimum fraction of galaxies that are likely to accrete above our luminosity threshold throughout the time elapsed in the redshift range probed. This will provide a corrected minimum active galaxy fraction within each redshift range.

\subsubsection{Corrected Active Fraction}

If we assume the cosmological principle and conclude that a snapshot of, for example, Fig. 6 were taken in a few hundred Myrs time it would look primarily unchanged, then, from the maximum timescales $\tau_{max}$, we must also conclude that many of the galaxies on the plot will no longer be visible as they must have ceased accreting at that rate so as to not end up higher than the local $M_{BH} - M_{*}$ relation. Therefore, we deduce the minimum fraction of galaxies that will at some point be visible above our luminosity limit (L$_{X}$ $>$ 2.35 $\times$ 10$^{43}$ erg s$^{-1}$) within the time allotted by a specific redshift range. This is a minimum value because we likely miss some active galaxies within our luminosity range due to very high obscuration along the line of sight for some galaxies as a result of torus opening angles aligned perpendicular to us, and further because we apply the inverse of the maximum lifetime of AGN as calculated in the previous section as a correction factor. We shall call the corrected minimum active fraction f$_{true}$, which is defined as:

\begin{equation}
f_{true}^{z1 - z2} = \frac{t_{z2} - t_{z1}} {< \tau_{max} >} f_{obs}^{z1 - z2}
\end{equation}

\noindent Where f$_{obs}$ is the observed AGN fraction as defined in \S 5.1, $< \tau_{max} >$ is the average value of $\tau_{max}$ (defined above) in the redshift range being considered, and t$_{z}$ is the age of the Universe at redshift z, such that $t_{z2} - t_{z1}$ is the time interval between redshifts $z1$ and $z2$.

The total fraction of galaxies expected to have X-ray luminosities above our threshold, therefore, varies with redshift range: At 0.4 $<$ z $<$ 1.0 it is 4.0 +/- 0.8 \%, at 1.0 $<$ z $<$ 1.5 it is 9.6 +/- 1.3 \% and at 1.5 $<$ z $<$ 3.0 it is 23.3 +/- 2.2 \%. These values may be taken as estimates of the total fraction of massive galaxies expected to reach bright (L$_{X}$ $>$ 2.35 $\times$ 10$^{43}$ erg s$^{-1}$) Seyfert level X-ray luminosities in each of the given redshift ranges for massive galaxies, with M$_{*}$ $>$ 10$^{10.5}$ M$_{\odot}$. Naturally, these values are in general higher than the direct measures of AGN fractions calculated in \S 5.1, because the maximum timescales are shorter than the length of time between the redshift limits of each bin. However, both measures of AGN fraction rise with redshift out to z $\sim$ 3.

This corrected fraction is plotted in Fig. 3 alongside the uncorrected AGN fraction evolution. The solid blue line represents the best simple power law fit to the data. Explicitly this fit is: f(L$_{X}$ $>$ 2.35 $\times$ 10$^{43}$ erg s$^{-1}$) = (1.16+/-0.4)(1 + z)$^{2.5+/-0.2}$.

\subsubsection{The Total Number of Active Galaxies since z = 3}

A question that has remained contentious, despite several conjectures and studies, is how many galaxies undergo an active phase in their development over the age of the Universe? It is clear that this question will depend on several factors, including the luminosity threshold at which one chooses to define a galaxy as being `active', the lifetime of AGN within this range, the stellar mass range of host galaxies considered, perhaps the environment in which the host galaxy resides, as well as numerous other (frequently ill defined) issues. We are in a position to place a minimum constraint on this question for $M_{*} > 10^{10.5} M_{\odot}$ galaxies (from the volume limited subset of the GNS and POWIR NIR surveys), and a definition of active as L$_{X}$ $>$ 2.35 $\times$ 10$^{43}$ erg s$^{-1}$ (from matching with the deepest available chandra X-ray data), since z = 3, utilising the maximum lifetimes and apparent active fraction evolution defined and calculated in previous sections (\S 5.1 and \S 5.4.1). This study will provide new limits on the total active fraction out to high redshifts, for `Seyfert' luminosity AGN in massive galaxies.

From our calculations of the average maximum lifetime of an AGN in our sample, and our fitting of the active fraction evolution, we calculate the fraction of $M_{*} > 10^{10.5} M_{\odot}$ galaxies that will be AGN (with L$_{X}$ $>$ 2.35 $\times$ 10$^{43}$ erg s$^{-1}$) since z = 3. We define the parameter, $\Gamma_{AGN}$, in analogy to the characteristic time between galaxy mergers (discussed in Bluck et al. 2009, Conselice, Yang \& Bluck 2009) which may be expressed as:

\begin{equation}
\Gamma_{AGN}(z) = \frac{< \tau_{max} >} {f_{AGN}(z)}
\end{equation}

\noindent where, $f_{AGN}$ is the observed fraction of AGN at redshift z ($f_{AGN} = 1.2(1 + z)^{1.6}$ from section 4.1), and $< \tau_{max} >$ $\sim$ 0.9 Gyr is the mean maximum lifetime of AGN over all of the redshifts being considered (i.e. z = 0 - 3), and we find no evidence of evolution of $\tau_{max}$ with redshift. Thus, $\Gamma_{AGN}$ must be interpreted as a maximum, with its inverse a minimum AGN rate. This is because the apparent active fraction is a minimum and the computed mean lifetime is a maximum. The significance of this parameter only becomes apparent (unlike $\Gamma$ for mergers) when we consider the time integral of its inverse. Here, the minimum total fraction of AGN ($F_{AGN}$) which will accrete with L$_{X}$ greater than some limit between z1 and z2, may be found by:

\begin{equation}
F_{AGN} = \int_{t1}^{t2} \Gamma_{AGN}^{-1}(z) dt = \int_{z1}^{z2} \Gamma_{AGN}^{-1}(z) \frac{t_{H}} {(1+z)} \frac{dz} {E(z)}
\end{equation}

\noindent where $\Gamma_{AGN}(z)$ is defined above, $t_{H}$ is the Hubble time, and $E(z) = [\Omega_{M}(1+z)^{3} + \Omega_{k}(1+z)^{2} + \Omega_{\Lambda}]^{1/2} = H^{-1}(z)$. Calculating this function from z = 0 to z = 3, utilising the power law parameterisation $f_{AGN} = 1.2(1 + z)^{1.4}$ and taking $< \tau_{max} >$ = 0.9 Gyr (see \S 5.4.1), we find that $F_{AGN}(z<3)$ = 0.41 +/- 0.06, implying that at least $\sim$ 40\% of all $M_{*} > 10^{10.5} M_{\odot}$ galaxies will have been AGN, accreting with hard band X-ray luminosities L$_{X}$ $>$ 2.35 $\times$ 10$^{43}$ erg s$^{-1}$, over the past $\sim$ 11.5 Gyrs (since z = 3). We note again here that this value is a minimum, and the liklihood is that the true fraction will be higher, perhaps much higher. However, this provides a lower limit to the fraction of massive galaxies which will have undergone an active phase (at bright `Seyfert' X-ray luminosities) since z = 3, and can be used to compute further results.

\subsubsection{The Energy Output of Active Galaxies since z = 3}

When considering the role of AGN in galaxy formation and evolution, a crucial quantity to know is the total energy released by AGN over their lifetimes. This quantity will impact on the feedback mechanisms involved in the co-evolution of SMBHs and their host galaxies, as well as contributing to our understanding of the X-ray background of the Universe. Almost all previous studies of this quantity have involved probing the evolution of the X-ray luminosity function (XLF) with redshift (e.g. Hasinger et al. 2005, Barger et al. 2005), or equivalently studying the comoving space density evolution of active galaxies with redshift (e.g. Ueda et al. 2003). These studies agree that there is a steep rise in the comoving X-ray luminosity density in the Universe with redshift out to z $\sim$ 1.5, with a less steep decline thereafter. This compares well with our results in \S 5.2, where we note a rise in $\mu M_{BH}$ ($\propto L_{X}$) with redshift out to z $\sim$ 1.5, with a possible flattening off thereafter.

In this section we compute the energy output due to sub-QSO AGN per massive ($M_{*} > 10^{10.5}$ M$_{\odot}$) galaxy in the Universe, by combining the maximum lifetimes of our sample of AGN, the minimum total active fraction of massive galaxies since z = 3, and the mean minimum bolometric luminosity of our AGN. Therefore, we can compute the energy released from AGN per massive galaxy as:

\begin{equation}
E_{AGN} = F_{AGN} \times < L_{Bol} > \times < \tau_{max} >
\end{equation}

\noindent where $F_{AGN}$ is the minimum total fraction of galaxies that will be active, with L$_{X}$ $>$ 2.35 $\times$ 10$^{43}$ erg s$^{-1}$, since z = 3, $< L_{Bol} >$ is the mean minimum bolometric luminosity of our sample of AGN, and $< \tau_{max} >$ is the mean maximum lifetime of our sample of AGN, which is $\sim$ 0.9 Gyr. Since the term $F_{AGN}$ contains a $< \tau_{max} >^{-1}$ term, the $< \tau_{max} >$ terms in $E_{AGN}$ cancel, yielding an estimate of the true total energy output due to AGN per massive galaxy in the Universe per mean bolometric luminosity. Thus, our value for E$_{AGN}$ is a lower limit to the total energy output due to AGN in massive galaxies, since we use a minimum bolometric correction. Furthermore, this must still be considered a minimum even if we were to use actual bolometric corrections as it does not take into account energy absorbed through Compton thick accretion discs, the host galaxy's interstellar medium (ISM), the local environment, and the intergalactic medium (IGM). Computing this, we find that $E_{AGN}$ = 1.4 +/- 0.25 $\times$ 10$^{61}$ erg per $M_{*} > 10^{10.5} M_{\odot}$ galaxy since z = 3. This corresponds to an average AGN luminosity density of $L_{AGN}$ = 1.0 +/- 0.3 $\times$ 10$^{57}$ erg Mpc$^{-3}$ Gyr$^{-1}$

Furthermore, using the Virial Theorem, we can define the binding energy of a galaxy ($V_{Gal}$) as:

\begin{equation}
V_{Gal} \sim M_{Gal} \times \sigma^{2}
\end{equation}

\noindent where $\sigma$ is the velocity dispersion of the galaxy, and $M_{Gal}$ is the total dynamical mass of the galaxy. Assuming an average velocity dispersion of 250 km/s and an average dynamical mass of $10^{11.5} M_{\odot}$ for our galaxies, we find a binding energy of V$_{Gal}$ = 3.95 $\times$ 10$^{59}$ erg. We define the ratio (r) between these energies such that:

\begin{equation}
r = \frac{E_{AGN}} {V_{Gal}} \sim 35
\end{equation}

\noindent Therefore, the total energy output due to all SMBH accretion since z = 3 is at least 35 times greater than the binding energy of all $M_{*} > 10^{10.5} M_{\odot}$ galaxies in the Universe. This minimum compares very favourably with prior theoretical and observational estimates of the energy emitted in forming a SMBH of $E_{SMBH} \sim 100 V_{Gal}$, found for example in Silk \& Rees (1998) and Fabian (1999).

\subsubsection{The Luminosity Density of Active Galaxies out to z = 3 - Explicit and Galaxy Methods}

We calculate the comoving X-ray luminosity density evolution for massive galaxies explicitly from the total X-ray luminosities emitted by massive galaxies in our surveys and the total volume to which we are complete. Specifically, we compute the X-ray luminosity density of AGN ($\rho_{AGN}$) by summing over the survey area and redshift range of our volume limited sample, which is explicity computed by:

\begin{equation}
\rho_{AGN}(z_{1} - z_{2}) = \frac{1}{V_{S}} \sum_{j = 0}^{j = A_{S}} \sum_{i = z1}^{i = z2} L_{X}(i,j)
\end{equation} 

\noindent where,

\begin{equation}
V_{S} = \frac{A_{S}}{4\pi} \left(\frac{\pi}{180}\right)^2 V_{C}
\end{equation}

\noindent where $V_{S}$ is the total comoving volume of our surveys to which we achieve mass and luminosity completeness at $M_{*} > 10^{10.5} M_{\odot}$ and $L_{X} > 2.35 \times 10^{43}$ erg s$^{-1}$. $A_{S}$ represents the fraction of each survey's area to which we are complete, and $V_{C}$ is the total comoving volume of the Universe between redshifts $z_{1}$ and $z_{2}$, computed for a flat spacetime in a $\Lambda$CDM cosmology, as defined in the introduction (\S 1).

We compute the evolution in the X-ray luminosity density with redshift, finding values of: 7.3 (+/-2.6) $\times$ 10$^{38}$ erg s$^{-1}$ Mpc$^{-3}$ at z = 0.4 - 1,  3.1 (+/-1.5) $\times$ 10$^{39}$ erg s$^{-1}$ Mpc$^{-3}$ at z = 1 - 1.5, and  5.4 (+/-2) $\times$ 10$^{39}$ erg s$^{-1}$ Mpc$^{-3}$ at z = 1.5 - 3. The lower two redshift ranges are in very good agreement with values computed from XLF studies for similar X-ray luminosity ranges to the ones we probe here, as found in Aird et al. (2009). This suggests that the vast majority of the energy contribution to the XLF comes from the most massive galaxies in the Universe at redshifts z = 0.4 - 1.5. These values are plotted against the total X-ray luminosity density of all sources in Fig. 10.

At z $>$ 1.5, however, we obtain a value for the X-ray luminosity density of massive galaxies that exceeds the total X-ray luminosity density computed for all galaxies across all X-ray luminosity ranges in Aird et al. (2009). This disagreement is most likely caused by the fact that the GNS (which we use exclusively to probe high redshifts, z $>$ 1.5, due to its high depth) is centred around massive galaxies at high redshifts, thus introducing a systematic bias to observe higher X-ray densities than average in the Universe due to us having higher than representative massive galaxy densities. We estimate our GNS field to be 1.6 times more dense in massive galaxies than the entire (not mass selected) GOODS field (Conselice et al. 2010; Mortlock et al. 2010 in prep.).

In principle this bias can be corrected for by computing the X-ray luminosity density for massive galaxies from a galaxy comoving number density approach, combining the energy output per galaxy and fraction of active galaxies (computed in \S 5.4.3 and 5.4.4) to the comoving number density evolution of massive galaxies from unbiased fields.

\begin{figure*}
\includegraphics[width=0.45\textwidth,height=0.45\textwidth]{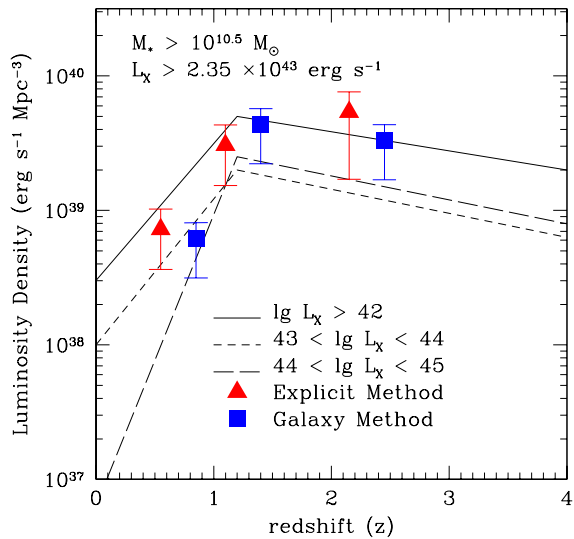}
\caption{The contribution to the X-ray luminosity function of sub-QSO massive galaxies across three redshift ranges, z = 0.4 - 1, z = 1 - 1.5, and z = 1.5 - 3. Our results are broadly consistent with X-ray active sub-QSO massive galaxies being the dominant source of X-ray radiation in the Universe, contributing the majority of the total X-ray luminosity density in each redshift range. Red triangles have luminosity densities computed via an explicit method computing the total X-ray luminosity within a given volume surveyed (see \S 5.4.5 for more details). Blue squares have luminosity densities computed via convolution with the comoving number densities of massive galaxies which agree (to within 1 $\sigma$) with the explicit method in the lower two redshift ranges, with more significant departure in the highest redshift range due to the fact that the GNS is selected to maximise the density of massive galaxies at z $>$ 1.7 in the GOODS field (see \S 5.4.6 for more details). The lines plot are approximate best fit X-ray luminosity functions to the deepest Chandra surveys (CDF-N/S) constructed from data in Aird et al. 2009, where we reproduce only the most relevant luminosity ranges here.}
\end{figure*}

From this approach, the average X-ray luminosity density, for the entire redshift range probed (z = 0.4 - 3) is 1.5 +/- 0.6 $\times$ 10$^{39}$ erg s$^{-1}$ Mpc$^{-3}$, but this varies with redshift as the active fraction and mean luminosity also vary. i.e the X-ray luminosity density at z = 0.4 - 1 it is 6.1 +/- 2.5 $\times$ 10$^{38}$ erg s$^{-1}$ Mpc$^{-3}$, at z = 1 - 1.5 it is 4.3 +/- 1.7 $\times$ 10$^{39}$ erg s$^{-1}$ Mpc$^{-3}$, and at z = 1.5 - 3 it is 3.3 +/- 1.2 $\times$ 10$^{39}$ erg s$^{-1}$ Mpc$^{-3}$. These values compare remarkably well to the calculation from the explicit method above for the lower two redshift ranges, and to calculations of the emissivity from XLF studies in Aird et al. (2009) in particular for all redshift ranges. These values are plotted in Fig. 10 alongside those calculated from the explicit method.

We find a rise in X-ray luminosity density with redshift out to z $\sim$ 1.5, followed by a levelling off, or probable turn around thereafter. Moreover, we find estimates of the emissivity for AGN with 43.5 $<$ log L$_{X}$ $<$ 45 with a peak value of $\sim$ 4 $\times$ 10$^{39}$ erg s$^{-1}$ Mpc$^{-3}$ in very close accord with Aird et al. (2009). It is pertinent to note that our estimates of this value are constructed from an independent methodology, and the fact that these different approaches agree is indicative of our ability to place definite precision constraints on the energy outpourings of AGN out to high redshifts, for modest luminosity Seyferts as well as QSOs on a galaxy population basis. Additionally, the corroboration between different methods obtaining comparable results lends further reliability to our estimates of the maximum timescales of AGN in our sample, and the minimum value of the total AGN fraction since z = 3. These results also suggest that massive galaxies are the dominant source of X-ray emission in the Universe across all of the redshift ranges probed, z = 0.4 - 3, since their contribution to the XLF is almost sufficient to account for its entire magnitude at each redshift.

\subsection{Hardness-Mass Dependence}

The hardness of an X-ray source can prove to be an interesting probe of various aspects of the nature of the AGN. In particular, it can reveal the level of absorption experienced and, hence, is coupled to the n(H) column density around AGN. We investigate here whether there are correlations between the hardness of our X-ray sources and the masses of the host galaxies in which these sources reside. We define hardness, H, as:

\begin{equation}
H = \frac{f_{H} - f_{S}} {f_{H} + f_{S}}
\end{equation}

\noindent where f$_{H}$ is the hard band flux and f$_{S}$ is the soft band flux. We crudely define hard X-ray sources as those with $H > 0.5$, and soft X-ray sources as those with $H < 0.5$. As we look to higher redshifts both the hard and soft bands will become effectively harder. However, for the purposes of this analysis, that effect is not a particularly great problem since at all rest-frame energies probed in this paper, softer bands will be more prone to absorption than harder bands, and, hence, this hardness ratio will still divide our sample amply between those with high and low levels of absorption. From Fig. 6 it is apparent that hard X-ray sources fall systematically below their softer counterparts in units of $\mu M_{BH}$. This might be expected as the harder X-ray sources must have experienced greater absorption due to higher n(H) column densities. This effect, however, is small and our results remain primarily unchanged even if we exclude the hard X-ray sources. We have minimised this effect by choosing to use the hard band X-ray data from Chandra where absorption is reduced.

An additional trend may also be noted whereby the harder sources tend to be more biased towards the high stellar mass end of these plots, in the lower two redshift ranges. In the highest redshift range selection effects conspire to lead us to detect predominantly hard X-ray sources and to our measure of hardness being skewed to harder sources generally. To test this we compare the hardness of X-ray sources with the stellar mass of their host galaxies for our lower redshift points.

We ran a KS test finding that the mean mass of host galaxies with a soft X-ray source was 8 $\times$ 10$^{10}$ M$_{\odot}$ +/- 1.5 $\times$ 10$^{10}$ M$_{\odot}$, with the mean mass of galaxies with a hard X-ray source being 1.1 $\times$ 10$^{11}$ M$_{\odot}$ +/- 1.5 $\times$ 10$^{10}$ M$_{\odot}$, across the lowest redshift ranges. This represents a 97 \% chance that the masses of the host galaxies for the hard and soft X-ray sources come from different underlying populations, a significance of $\sim$ 2 $\sigma$.

\section{Discussion}

In this section we give a summary of our results and discuss them in the context of galaxy formation and evolution, and in particular the role of AGN within the evolution of galaxies at $z < 3$.

\subsection{Evolution in the $M_{BH} - M_{*}$ Relationship}

It is difficult to disentangle the SMBH mass from measures of the Eddington limiting (minimum) SMBH mass, without knowing something about the Eddington ratio, $\mu$. As such, we have retained this source of error within our data and chosen to plot $\mu M_{BH}$ (= $M_{E}$) as opposed to $M_{BH}$ in all plots, making no assumption about the actual value of $\mu$, and instead representing this degeneracy by showing the relative position of the local $M_{BH} - M_{*}$ relation as a function of $\mu$ (solid and dotted lines) in Figs. 6 and 7.  We have noted some evolution in the $\mu M_{BH} / M_{*}$ ratio with redshift, signifying either a possible departure from the local relationship at higher redshifts, or evolution in the mean value of $\mu$. If we assume that the local relation remains valid at high redshifts then this corresponds to an observed evolution in $\mu$. 

The fact that we observe a higher fraction of AGN with $\mu$ $>$ 0.1, and a higher mean value of $\mu$, at higher redshifts is probably indicative of the amount of cool gas available in the centre of galaxies decreasing over cosmic time, since the $\mu$ ratio is directly related to the available fuel for the SMBH. Where there is unlimited fuel in the form of cool gas at the centre of galaxies, and excellent supply routes to get this fuel into the black hole, one would expect the accretion luminosity to tend towards the Eddington limiting luminosity, giving $\mu$ = 1. With a reduction in fuel or supply route efficiency (e.g. through obstruction from outflows) one would expect a lowering in the Eddington ratio. Since galaxies in general decrease their cold gas content over cosmic time due to supernovae (SNe) and AGN feedback, merging, and environmental effects (such as tidal and ram pressure stripping), it is natural to expect a reduction in the mean value of $\mu$ over cosmic time (see respectively Brighenti \& Mathews 2003, Ciotti \& Ostriker 2001, Bluck et al. 2009, Conselice, Yang \& Bluck 2009, and Roediger \& Hensler 2005).

If evolution in $\mu$ exists then there is no requirement for further evolution in the $M_{BH} - M_{*}$ relation. Thus, our results are consistent with the view that SMBHs and their host galaxies largely co-evolve, increasing their mass in proportion to one another. However, if we take the converse view that there is no evolution in $\mu$ then we must conclude that the $M_{BH} - M_{*}$ relation evolves such that SMBHs had masses higher than expected from the local relation, with regards to their host galaxy's mass, at higher redshifts. Given that the mean value of $\mu$ is expected to rise with redshift in general for massive galaxies, we can quantify the maximum possible evolution in the local  $M_{BH} - M_{*}$ relation by setting $\mu$ to be constant at its low redshift value and permitting only evolution in $M_{BH} - M_{*}$ to account for the observed evolution in $\mu M_{BH} - M_{*}$. Quantitatively this corresponds to a maximum evolution in the ratio $M_{*} / M_{BH}$ of less than a factor of two, from $\sim$ 1000 at z = 0 to $\sim$ 700 at z = 2.5. 

This type of evolution to lower $M_{*} / M_{BH}$ values at higher redshifts seems unlikely as it would imply either that SMBHs lose mass (which is ruled out on theoretical grounds) or else that galaxies grow in size more rapidly than their central black holes. This is also very unlikely as we have observed a sharp decline in merging and star formation rate with redshift (see e.g. Bauer et al. 2010 in prep., Bluck et al. 2009, Conselice, Yang \& Bluck 2009) for massive galaxies at redshifts z $<$ 3, and moreover SMBHs are expected to grow due to merging in line with their host galaxies. Therefore, we prefer to interpret our results as evolution with redshift of the Eddington ratio, $\mu$, further implying that the available fuel for SMBHs declines over cosmic time, which is explicable within the current paradigm of galaxy evolution. As galaxies evolve they experience feedback from AGN and SNe which both remove the availability of cold gas, and furthermore merging and environmental effects may also contribute to this reduction in cool gas content of galaxies. As such, older galaxies have less available fuel at their centres for their SMBHs and, hence, they accrete at lower fractions ($\mu$) of the Eddington maximum.

One other possibility is that the evolution is $\mu$ is dramatic, and thus the local $M_{BH} - M_{*}$ relationship evolves such that SMBHs were {\it less} massive than expected for their host galaxies' masses. We can quantify this maximum positive evolution in the $M_{BH} / M_{*}$ ratio as being $\sim$ a factor of 10, by setting $\mu$ = 1, its maximum limit assuming sub-Eddington accretion. Thus, we conclude that either SMBHs evolve more or less in proportion to their host galaxies, or else there is dramatic evolution in $\mu$ leading to SMBHs growing later in mass than their host galaxies, through accretion of cool gas.

By way of comparison to the body of literature on the subject of possible evolution in the $M_{BH} - M_{*}$ relationship with redshift, it is pertinent to note that a wide variety of methods and luminosity ranges have been used and studied, leading to a quite diverse set of conclusions. For example, Borys et al. (2005) find that SCUBA galaxies have SMBHs which are systematically lower than expected from the local relation for their stellar mass by up to a factor of 50 or so (via Eddington methods), and Alexander et al. (2008) also deduce that for SMGs (sub millimetre galaxies) SMBHs are also smaller than expected for their host galaxy's mass by a smaller factor of 3 or so (via virial methods). However, Jahnke et al. (2009) find that type-1 AGN galaxies from the COSMOS survey fit closely onto the local $M_{BH} - M_{*}$ slope of Haring \& Rix (2004) out to z $\sim$ 2. Conversely, recent studies by Merloni et al. (2010) and Declari et al. (2010a,b) using virial estimators of SMBH masses in QSO's both conclude that there is evolution in the local $M_{BH} - M_{*}$ relation with redshift such that $M_{*} / M_{BH}$ was lower at high z, by up to a factor of 7 (Declari et al. 2010) or 3 (Merloni et al 2010) by z = 3.

Our results are for lower luminosity AGN, principally bright Seyferts, than the majority of studies discussed above, and we rule out positive evolution in the $M_{*} / M_{BH}$ ratio of greater than a factor of 2, but we do not particularly well constrain evolution to higher values (noting that this must be less than a factor of $\sim$ 10). Despite the inconsistency in luminosity ranges probed and methods used for deducing SMBH masses, there is very broad agreement between all of these approaches that the local $M_{BH} - M_{*}$ relationship is not departed from by more than roughly one order of magnitude (or a factor of 10) in either direction by z = 3. Our study particularly favours a close adherence to the local relationship of Haring \& Rix (2004) for bright Seyfert galaxies out to z = 3, providing there is not drastic positive evolution in the Eddington ratio, $\mu$, with redshift. If this type of drastic increase in $\mu$ does occur then we are forced to conclude that SMBHs in Seyfert galaxies form their mass after their host galaxies, in apparent contradiction to the results for QSO's in Merloni et al. (2010) and Declari et al. (2010a,b), but in accord with the conclusions based on studying SMGs in Borys et al. (2005) and Alexander et al. (2008). Nevertheless, we feel that the most natural conclusion from our results is that SMBHs and their host galaxies evolve principally together, in proportion to one another, preserving the local relation out to z = 3, in agreement with Jahnke et al. (2009), as this would require only mild and plausible evolution in $\mu$ with redshift.

\subsection{From AGN Lifetimes to Feedback Energy}

We observe that the fraction of AGN above our luminosity threshold ($L_{X}$ $>$ 2.35 $\times$ 10$^{43}$ erg s$^{-1}$) rises with redshift, and that the true fraction, calculated via considering the maximum amount of time a given AGN can continue accreting at its current rate before surpassing the local relation, also rises with redshift. This result reflects the rising mean Eddington ratio with redshift, assuming the validity of the local $M_{BH} - M_{*}$ relation at all redshifts. Together this may all be interpreted as AGN activity in the Universe rising with redshift, and possibly levelling out at z $\sim$ 1.5 - 3 where there is an optimum synergy between the mass of SMBHs and available fuel at the centre of each galaxy. This is consistent with conclusions based on probing the X-ray luminosity function (XLF), in e.g. Ueda et al. (2003), Barger et al. (2005), Hasinger et al. (2005) and Aird et al. (2009), who all find a similar rise and then peak in X-ray luminosity within a similar redshift range.

We have constructed a novel method for estimating the maximum lifetimes of Seyfert luminosity AGN. Using the timescales calculated from considering the maximum accretion permitted before a given AGN will rise above the local $M_{BH} - M_{*}$ relation, we have calculated the total fraction of $M_{*} > 10^{10.5} M_{\odot}$ galaxies which will be active, with $L_{X}$ $>$ 2.35 $\times$ 10$^{43}$ erg s$^{-1}$, since z = 3. We find that at least 40\% of all massive galaxies will reach Seyfert level luminosities from accretion onto their central SMBHs. Perhaps all massive galaxies will undergo an active period in their evolution since z = 3, but currently we can only be sure that at least a significant fraction will obtain this level of X-ray activity.

Moreover, we have computed the energy density due to seyfert luminosity AGN residing within massive galaxies in the Universe as $\rho_{AGN}$ = 1.0 +/- 2.5 $\times$ 10$^{57}$ erg Mpc$^{-3}$ Gyr$^{-1}$, noting that this corresponds to an average massive galaxy producing at least a factor of $\sim$ 35 times its binding energy in AGN emission. Further, we compute the X-ray luminosity density from this method and note that it rises with redshift from 6.1 +/- 2.5 $\times$ 10$^{38}$ erg s$^{-1}$ Mpc$^{-3}$ at z = 0.4 - 1 to 4.3 +/- 1.7 $\times$ 10$^{39}$ erg s$^{-1}$ Mpc$^{-3}$ at z = 1 - 1.5, with a slight reduction to 3.3 +/- 1.2 $\times$ 10$^{38}$ erg s$^{-1}$ Mpc$^{-3}$ at z = 1.5 - 3, in very close accord to other, alternative, XLF based methods, including Aird et al (2009). In fact in Fig. 10 we compare the total X-ray luminosity density in the Universe at various redshifts (found in Aird et al. 2009) to the contribution provided by sub-QSO massive Seyfert luminosity galaxies (this paper), finding that the majority of the X-ray luminosity density of the Universe is provided by this population at all redshifts up to z = 3.

It has been known for some time that the energy outpourings of AGN will be greater than the binding energy of their host galaxies (see e.g. Silk \& Rees 1998, Fabian et al. 1999), but in this paper we have constrained this empirically per galaxy, as opposed to per AGN. We now know that not only will active galaxies have a greater energy output by their AGN than their total binding energy by some large factor (up to 100 fold according to Fabian et al. 1999), but that the energy released by AGN (even primarily in the Seyfert regime) is at least 35 times greater than the binding energy of all massive ($M_{*} > 10^{10.5} M_{\odot}$) galaxies in the Universe. However, it is clear that the crucial question still to be answered is, what fraction of this energy couples to the galaxy? What we measure is the energy that escapes the galaxy to be later detected by us, whereas what matters for considerations of AGN feedback on star formation is what magnitude of energy couples to the galaxy itself through radiative or momentum driven processes. To improve these minima, apply them to average mass galaxies, and develop a deeper understanding of the coupling of AGN emissions to galaxy evolution, we shall have to wait for next generation X-ray telescopes and surveys, as well as the JWST.

\section{Summary and Conclusions}

In this paper we have detected 508 X-ray selected AGN at 0.4 $<$ z $<$ 6 across the EGS and GOODS North and South fields, using the criteria set out in \S 3.1. We have constructed a volume limited sample of 85 galaxies with $L_{X}$ $>$ 2.35 $\times$ 10$^{43}$ erg s$^{-1}$ and M$_{*}$ $>$ 10$^{10.5}$ M$_{\odot}$ at 0.4 $<$ z $<$ 3, with which we probe the co-evolution of SMBHs and their host galaxies over the past 11.5 billion years. We adopt an Eddington limit method to obtain minimum mass estimates for the SMBHs residing at the centre of these active galaxies and compare these values to stellar masses of the host galaxies as a function of redshift (see \S A for data table). We find an active fraction of 1.2 +/- 0.2 \% at 0.4 $<$ z $<$ 1, which rises to 6.0 +/- 0.8 \% at 1 $<$ z $<$ 1.5, with a further modest rise to 7.4 +/- 2.0 \% at 1.5 $<$ z $<$ 3.

We probe the $\mu M_{BH} - M_{*}$ relation across three redshift ranges (see Fig. 6) and find that this relation appears in general lower in value, and less steep, than the local $M_{BH} - M_{*}$ relation. We interpret these differences as lying with our inability to detect low mass host galaxies accreting with small values of $\mu$ and, further, interpret the offset between the local $M_{BH} - M_{*}$ relation and our measured $\mu M_{BH} - M_{*}$ relation as an evolution in $\mu$. In particular, we measure the fraction of high accreting SMBHs, accreting with $\mu$ $>$ 0.1 (assuming the local relation's validity at z $>$ 0), to rise from $\sim$ 17\% at z = 0.6 to $\sim$ 22\% at z $\sim$ 2.5 (representing a 3$\sigma$ evolution).

We calculate accretion rates for our sample of AGN and plot the results of this in Fig. 9. We find that there is a high degree of scatter in the accretion rate across all redshifts and stellar masses. Nonetheless, we identify a population of galaxies with very high accretion rates which are only found in high stellar mass galaxies. Moreover, we find that higher mass host galaxies contain higher accreting AGN to a 3$\sigma$ significance. This is to be expected because more massive galaxies are predicted to contain more massive SMBHs which may accrete at a higher rate.

Further, we calculate maximum timescales at which AGN in our sample can continue to accrete at their current rate, noting that many of these timescales are quite short, with maximum values $\sim$ a few 100 Myrs, and an expectation value of $< \tau_{max} >$ $\sim$ 0.9 Gyr. From this we calculate the true corrected fraction of `Seyfert' luminosity AGN in each redshift range, observing that it also rises with redshift.

Perhaps our largest contribution in this paper is computing a minimum fraction of galaxies at $M_{*} > 10^{10.5} M_{\odot}$ that will contain `Seyfert' luminosity AGN at some point over the past 11.5 Gyrs (since z = 3). We find that at least $\sim$ 40\% of all massive galaxies will at some point during their evolution between z = 3 to the present be active with $L_{X}$ $>$ 2.35 $\times$ 10$^{43}$ erg s$^{-1}$. We also compute the X-ray luminosity density evolution due to massive galaxies, with Seyfert luminosities and brighter, from our sample of AGN, finding a close agreement with the latest estimates of the total X-ray luminosity density evolution from distinct X-ray luminosity based methods (e.g. Aird et al. 2009). This suggests that the majority of the X-ray luminosity function is generated by massive sub-QSO galaxies. 

Moreover, we use this total active fraction at `Seyfert' luminosities to compute the average energy output due to AGN per galaxy, and note that this is at least 35 times greater than the binding energy of an average massive galaxy. This is consistent with several previous studies in the literature (e.g. Silk \& Rees 1998, Fabian 1999). We extend this prior work quoting this minima empirically and for each massive galaxy in the Universe, as opposed to each AGN. The impact of this much energy being released into the galaxies, as well as into the surrounding large scale environments, might have a profound influence on galaxies' formation and evolution through AGN feedback. But a crucial question remains: how much of this energy couples to the host galaxy through radiative or momentum driven processes, actually causing the feedback on star formation?

In conclusion, we have provided evidence for the necessity of redshift evolution in $\mu$ to account for our data trends: $\mu$ must evolve with redshift to higher values, indicating that galaxies at higher redshifts were more rich in the cold gas used as fuel by the SMBH. Furthermore, we note an increase in the AGN fraction with redshift which is also indicative of overall AGN activity rising with redshift, as seen previously in many studies of the evolution of the X-ray luminosity function (e.g. Barger et al. 2005, Hasinger et al. 2005). Thus, there is a rise in AGN luminosity out to z $\sim$ 1.5 - 2, with a possible levelling off thereafter to z = 3, driven perhaps by an optimal balancing of available fuel and mass of SMBHs. Our results are consistent with no change in the local $M_{BH} - M_{*}$ relation, providing there is modest and plausible evolution with redshift in $\mu$. Furthermore, we place a maximum limit on possible evolution in the $M_{*} / M_{BH}$ ratio of less than a factor of two to lower values, and less than a factor of ten to higher values, for massive galaxy `Seyferts' luminosity AGN. Therefore, we suggest that SMBH masses and their host galaxy stellar masses evolve together, in direct proportion to one another, or else there is a dramatic increase in the Eddington ratio with redshift and, thus, SMBH masses are built up after their host galaxies.

We thank Olivia Johnson for providing us with useful codes to simulate the effect of obscuration across hard to soft X-ray bands. We appreciate helpful discussions and comments from Amanda Bauer, Fernando Buitrago, Samantha Penny, Robert Chuter, Mark Dickinson, Alice Mortlock and Mat Page. We thank the anonymous referee for several insightful comments and suggestions on this work. We also gratefully acknowledge funding from the STFC to carry out this research.

\onecolumn

\appendix
\section{Data Tables}

\begin{longtable}{@{}cccccccc}
\caption{Volume Limited Sample of Active Massive Galaxies at 0.4 $<$ z $<$ 3}
\label{tab3} \\
\hline
\hline
\multicolumn{1}{c}{ID}&
\multicolumn{1}{c}{RA}&
\multicolumn{1}{c}{Dec}&
\multicolumn{1}{c}{z$_{phot}$}&
\multicolumn{1}{c}{Log($M_{*}$ [$M_{\odot}$])}&
\multicolumn{1}{c}{Log($L_{\rm {2-10KeV}}$ [erg s$^{-1}$])}&
\multicolumn{1}{c}{Log($M_{Edd}$ [$M_{\odot}$])}&
\multicolumn{1}{c}{Hardness (H)} \\
\hline
\endfirsthead

\multicolumn{7}{l}{{\tablename} \thetable{} --Continued}\\
\hline
\hline
\multicolumn{1}{c}{ID}&
\multicolumn{1}{c}{RA}&
\multicolumn{1}{c}{Dec}&
\multicolumn{1}{c}{z$_{phot}$}&
\multicolumn{1}{c}{Log($M_{*}$ [$M_{\odot}$])}&
\multicolumn{1}{c}{Log($L_{\rm {2-10KeV}}$ [erg s$^{-1}$])}&
\multicolumn{1}{c}{Log($M_{Edd}$ [$M_{\odot}$])}&
\multicolumn{1}{c}{Hardness (H)} \\
\hline
\endhead

 1 & 189.1483 &  62.2400 & 1.550 & 10.84 & 43.42 & 6.42 & 0.85 \\ 
 2 & 189.0489 &  62.1708 & 1.550 & 10.72 & 43.46 & 6.46 & 0.95 \\ 
 3 & 189.0659 &  62.2544 & 2.745 & 11.25 & 43.85 & 6.85 & 0.95 \\ 
 4 & 189.0912 &  62.2677 & 2.235 & 10.75 & 43.51 & 6.51 & 1.00 \\ 
 5 & 189.0798 &  62.2449 & 2.945 & 10.76 & 43.95 & 6.95 & 0.28 \\ 
 6 & 189.0944 &  62.1747 & 2.385 & 10.61 & 43.91 & 6.91 & 0.30 \\ 
 7 &  53.1070 & -27.7182 & 2.410 & 11.27 & 44.17 & 7.17 & 0.59 \\ 
 8 &  53.0764 & -27.8486 & 1.570 & 11.42 & 43.74 & 6.74 & 0.93 \\ 
 9 &  53.0393 & -27.8018 & 2.705 & 11.37 & 44.34 & 7.34 & 0.67 \\ 
10 & 213.8910 &  52.0994 & 0.986 & 11.06 & 44.18 & 7.18 & 0.17 \\ 
11 & 213.9490 &  52.1455 & 0.619 & 11.33 & 43.44 & 6.44 & 0.77 \\ 
12 & 213.7360 &  52.1612 & 0.992 & 11.32 & 43.77 & 6.76 & 0.42 \\ 
13 & 213.9640 &  52.2279 & 0.752 & 10.95 & 43.43 & 6.42 & 0.45 \\ 
14 & 214.2080 &  52.3026 & 0.808 & 11.17 & 43.41 & 6.41 & 0.97 \\ 
15 & 214.2220 &  52.3511 & 0.982 & 11.06 & 43.37 & 6.37 & 0.16 \\ 
16 & 214.4410 &  52.5090 & 0.985 & 11.34 & 43.85 & 6.85 & 0.70 \\ 
17 & 214.3860 &  52.5342 & 0.986 & 11.34 & 43.59 & 6.59 & 0.48 \\ 
18 & 214.4790 &  52.6956 & 0.464 & 10.52 & 43.82 & 6.82 & 0.85 \\ 
19 & 214.9280 &  52.7771 & 0.784 & 10.92 & 43.73 & 6.73 & 0.35 \\ 
20 & 214.7950 &  52.7839 & 0.820 & 11.15 & 43.60 & 6.59 & 0.51 \\ 
21 & 215.1170 &  52.9782 & 0.871 & 10.62 & 43.56 & 6.55 & 0.37 \\ 
22 & 213.7870 &  52.1117 & 0.837 & 11.09 & 43.65 & 6.65 & 0.54 \\ 
23 & 213.7930 &  52.1086 & 0.931 & 10.85 & 43.60 & 6.60 & 0.42 \\ 
24 & 214.2680 &  52.3611 & 0.859 & 11.17 & 43.77 & 6.77 & 0.89 \\ 
25 & 214.2140 &  52.3462 & 0.981 & 11.35 & 43.92 & 6.92 & 0.15 \\ 
26 & 214.0950 &  52.3212 & 0.826 & 11.35 & 44.47 & 7.47 & 0.42 \\ 
27 & 214.1440 &  52.3775 & 0.885 & 10.88 & 43.56 & 6.56 & 0.50 \\ 
28 & 214.4960 &  52.5275 & 0.938 & 11.02 & 43.46 & 6.46 & 0.51 \\ 
29 & 215.5220 &  53.1810 & 0.995 & 10.88 & 43.66 & 6.65 & 0.46 \\ 
30 & 215.5000 &  53.2119 & 0.780 & 11.51 & 43.92 & 6.91 & 0.51 \\ 
31 & 215.2640 &  53.3062 & 0.873 & 10.96 & 43.42 & 6.42 & 0.20 \\ 
32 & 215.7650 &  53.4468 & 0.866 & 10.50 & 43.66 & 6.66 & 0.46 \\ 
33 & 213.8120 &  52.1602 & 1.296 & 11.33 & 43.61 & 6.61 & 0.31 \\ 
34 & 214.2120 &  52.2579 & 1.034 & 10.89 & 44.24 & 7.23 & 0.35 \\ 
35 & 214.1370 &  52.3172 & 1.028 & 11.12 & 44.15 & 7.15 & 0.97 \\ 
36 & 214.0270 &  52.4165 & 1.313 & 11.37 & 44.34 & 7.34 & 0.60 \\ 
37 & 214.4240 &  52.4732 & 1.148 & 11.93 & 44.37 & 7.37 & 0.53 \\ 
38 & 214.3910 &  52.5637 & 1.083 & 11.43 & 43.40 & 6.40 & 0.66 \\ 
39 & 214.5230 &  52.6757 & 1.413 & 10.79 & 44.08 & 7.07 & 0.54 \\ 
40 & 214.9170 &  52.8274 & 1.232 & 11.32 & 43.42 & 6.42 & 0.69 \\ 
41 & 214.7870 &  52.9436 & 1.392 & 10.69 & 44.22 & 7.22 & 0.40 \\ 
42 & 215.4250 &  53.1798 & 1.418 & 11.26 & 43.60 & 6.59 & 0.30 \\ 
43 & 213.9240 &  52.1551 & 1.343 & 10.51 & 43.46 & 6.45 & 0.45 \\ 
44 & 213.6690 &  52.1407 & 1.017 & 11.02 & 43.59 & 6.59 & 0.50 \\ 
45 & 214.1800 &  52.2433 & 1.069 & 11.54 & 45.03 & 8.03 & 0.35 \\ 
46 & 214.2360 &  52.2577 & 1.033 & 11.49 & 43.44 & 6.44 & 0.19 \\ 
47 & 213.9910 &  52.2502 & 1.060 & 10.81 & 43.39 & 6.39 & 1.00 \\ 
48 & 213.9520 &  52.2691 & 1.033 & 10.93 & 43.91 & 6.90 & 0.44 \\ 
49 & 214.1500 &  52.3201 & 1.180 & 10.98 & 43.38 & 6.38 & 0.50 \\ 
50 & 214.4120 &  52.3925 & 1.138 & 10.68 & 43.91 & 6.90 & 0.35 \\ 
51 & 213.9730 &  52.3799 & 1.405 & 10.70 & 43.47 & 6.47 & 0.58 \\ 
52 & 214.1240 &  52.3926 & 1.044 & 11.49 & 43.47 & 6.47 & 0.97 \\ 
53 & 214.5000 &  52.3732 & 1.240 & 11.09 & 43.99 & 6.99 & 0.69 \\ 
54 & 214.4550 &  52.4699 & 1.064 & 11.07 & 43.62 & 6.62 & 0.62 \\ 
55 & 214.5860 &  52.5842 & 1.488 & 10.87 & 43.43 & 6.43 & 1.00 \\ 
56 & 214.5190 &  52.6092 & 1.223 & 11.15 & 43.88 & 6.88 & 0.30 \\ 
57 & 214.6920 &  52.6866 & 1.492 & 10.90 & 43.90 & 6.90 & 0.39 \\ 
58 & 214.5590 &  52.7224 & 1.206 & 10.88 & 43.55 & 6.55 & 0.90 \\ 
59 & 214.5770 &  52.7322 & 1.148 & 10.54 & 44.00 & 7.00 & 0.36 \\ 
60 & 214.4680 &  52.7154 & 1.239 & 10.91 & 43.52 & 6.52 & 0.69 \\ 
61 & 214.7520 &  52.7618 & 1.460 & 10.65 & 43.42 & 6.42 & 0.75 \\ 
62 & 215.0280 &  52.8236 & 1.443 & 10.67 & 43.64 & 6.63 & 0.21 \\ 
63 & 214.9480 &  52.8407 & 1.336 & 11.45 & 43.40 & 6.39 & 1.00 \\ 
64 & 215.2690 &  52.9748 & 1.322 & 11.03 & 43.77 & 6.77 & 0.88 \\ 
65 & 215.2820 &  53.0550 & 1.052 & 10.70 & 43.74 & 6.73 & 0.52 \\ 
66 & 214.9760 &  53.0463 & 1.419 & 10.97 & 43.70 & 6.70 & 0.28 \\ 
67 & 214.9720 &  53.0349 & 1.362 & 10.83 & 44.15 & 7.14 & 0.45 \\ 
68 & 215.0970 &  53.0865 & 1.078 & 11.43 & 43.55 & 6.55 & 0.89 \\ 
69 & 215.0850 &  53.1205 & 1.309 & 11.08 & 43.58 & 6.58 & 0.35 \\ 
70 & 215.4300 &  53.1478 & 1.200 & 10.95 & 43.61 & 6.61 & 0.59 \\ 
71 & 215.4930 &  53.1347 & 1.043 & 10.77 & 43.47 & 6.47 & 0.46 \\ 
72 & 215.0710 &  53.1338 & 1.038 & 11.18 & 43.68 & 6.68 & 0.34 \\ 
73 & 215.5000 &  53.2149 & 1.309 & 10.62 & 43.56 & 6.56 & 0.69 \\ 
74 & 215.3970 &  53.2262 & 1.150 & 11.17 & 43.74 & 6.73 & 0.33 \\ 
75 & 215.2950 &  53.1810 & 1.206 & 10.81 & 43.71 & 6.70 & 0.45 \\ 
76 & 215.4800 &  53.2373 & 1.226 & 10.83 & 43.80 & 6.80 & 0.38 \\ 
77 & 215.7630 &  53.3735 & 1.405 & 11.44 & 44.27 & 7.27 & 0.67 \\ 
78 & 215.6570 &  53.3620 & 1.272 & 11.05 & 43.96 & 6.96 & 0.43 \\ 
79 & 215.5830 &  53.3306 & 1.240 & 11.13 & 43.71 & 6.71 & 0.43 \\ 
80 & 215.4020 &  53.3373 & 1.466 & 11.08 & 43.88 & 6.88 & 0.18 \\ 
81 & 215.5160 &  53.4142 & 1.082 & 11.13 & 44.11 & 7.11 & 0.37 \\ 
82 & 215.8490 &  53.4718 & 1.460 & 10.90 & 44.19 & 7.18 & 0.91 \\ 
83 & 215.7700 &  53.5167 & 1.281 & 10.67 & 43.41 & 6.40 & 0.20 \\ 
84 & 215.7770 &  53.5192 & 1.281 & 10.69 & 44.22 & 7.21 & 0.91 \\ 
85 & 215.6980 &  53.5342 & 1.157 & 10.78 & 43.56 & 6.56 & 0.63 \\  
\hline
\end{longtable}

In the above table we restrict the full AGN matched list of 508 secure AGN detections (provided by matching within a 1.5$''$ radius GOODS and EGS galaxy catalogs with the extensive deep Chandra imaging in the same fields), by highlighting only those AGN with host galaxy stellar masses $M_{*} > 10^{10.5} M_{\odot}$ and with hard band X-ray luminosities $L_{\rm {2-10KeV}} > 2.35 \times 10^{43}$ erg s$^{-1}$, where we are complete for restricted areas of the EGS sample to z = 1.5, and for the GNS sub-sample to z = 3. $M_{*}$ is the stellar mass of the AGN host galaxy, $M_{Edd}$ is the Eddington limitted (minimum) mass of the SMBH (see \S 4.3 for definition), and H is the hardness ratio (see \S 5.5 for definition).

\twocolumn

\end{document}